\documentclass[11pt]{article}

\usepackage[usenames,dvipsnames,table]{xcolor}
\usepackage[utf8]{inputenc}

\pdfoutput=1 
\usepackage{jheppub} 

\usepackage{graphicx}
\usepackage{amsmath, amssymb}

\usepackage{mathrsfs}

\usepackage{framed}

\newcommand{\be}{\begin{eqnarray}}
\newcommand{\ee}{\end{eqnarray}}

 \definecolor{shadecolor}{rgb}{0.90,0.90,0.90}

\graphicspath{{./figures/}}

\def\KK{{\mathcal K}}
\def\cN{ {\mathcal N}}

\newcommand{\bea}{\begin{eqnarray}}
\newcommand{\eea}{\end{eqnarray}}  
\newcommand{\nn}{\nonumber}

\newcommand{\NN}{\mathcal{N}}

 \newcommand{\SU}{\mathrm{SU}}
 \newcommand{\SO}{\mathrm{SO}}
 
 \newcommand{\U}{\mathrm{U}}

 \newcommand{\vl}{\vec{\lambda} }
 
 \newcommand{\cO}{{\mathcal O}  }
 \newcommand{\bx}{ {\bf x}}
 \newcommand{\bu}{ {\bf u}}

\def\cO{ {\cal O}}
\def\cI{ {\cal I}}

\newcommand{\qPoc}[2]{\left(#1;#2\right)_\infty}
\newcommand{\qPocn}[3]{\left(#1;#2\right)_{#3}}
\newcommand{\GF}[1]{\Gamma\left(#1\right)}
\newcommand{\thf}[1]{\theta_p\left(#1\right)}

\newcommand{\thfq}[1]{\theta_q\left(#1\right)}
\newcommand{\thfQ}[1]{\theta_Q\left(#1\right)}
\newcommand{\pq}[2]{(pq)^{\frac{#1}{#2}}}

\title{On Ruijsenaars-Schneider spectrum from superconformal indices and ramified instantons}

\preprint{}

\author[a,b]{Hee-Cheol Kim,}

\author[c]{Anton Nedelin,}

\author[d]{Shlomo S. Razamat }

\affiliation[a]{Department of Physics, POSTECH, Pohang 37673, Korea}

\affiliation[b]{Jefferson Physical Laboratory, Harvard University, Cambridge, MA 02138, USA}

\affiliation[c]{Department of Mathematics, King’s College London,
London, WC2R 2LS, United Kingdom}

\affiliation[d]{Department of Physics, Technion, Haifa, 32000, Israel}

\emailAdd{heecheol@postech.ac.kr, anton.nedelin@gmail.com, razamat@physics.technion.ac.il}

\abstract{We discuss two physics-inspired approaches to derivation of the eigenfunctions and eigenvalues of 
$A_N$ Ruijsenaars-Schneider model. First approach which was recently proposed by the authors relies on the computations 
of superconformal indices of class $\mathcal{S}$ $4d$ $\NN=2$ theories with the insertion of surface defects.
Second approach uses computations of Nekrasov-Shatashvili limit of $5d$ $\NN = 1^*$ instanton partition functions in the 
presence of co-dimension two defect. We compare 
results of these two approaches for the low-lying levels of Ruijsenaars-Schneider model. We also discuss
different previously proposed exact quantization conditions for the Coulomb branch parameters of the instanton partition functions 
and their interpretations in terms of index calculations. }

\begin{document}

\maketitle
\flushbottom

\section{Introduction}
\label{sec:intro}

Supersymmetric quantum field theories in different dimensions are a fruitful setup to derive and test a plethora of results in mathematical physics. In particular it is common that the same mathematical structure appears in different physical setups allowing to perform some computations in different 
ways leading on one hand to deeper insights of the underlying physics and on the other hand allowing for derivation of mathematical results.

In this  note we discuss an example of such a mathematical structure, the $A_{N-1}$ Ruijsenaars-Schneider (RS) quantum mechanical integrable model. This model appears in numerous contexts in mathematical physics \cite{Ruijsenaars:1986pp,Ruijsenaars:1986vq,Ruijsenaars:1988pv,Gorsky:1993pe,Gorsky:1993dq,Gorsky:1994dj,Fock:1999ae,Braden:1996he,Braden:1997nc}. As with any quantum mechanical system an important question is that of spectrum of eigenfunctions
and eigenvalues of the model. For example, in the context of $4d$ class $\mathcal{S}$ theories \cite{Gaiotto:2009we,Gaiotto:2009hg} knowing this
spectrum conjecturally \cite{Gaiotto:2012xa,Gadde:2011uv} would allow
full determination of the superconformal index \cite{Kinney:2005ej,Romelsberger:2005eg,Dolan:2008qi} (which encodes a lot of information about protected operators) of theories for which  a Lagrangian description is not 
known, at least as of yet. Recently, an algorithmic procedure to deduce the  spectrum of RS model using the index of $4d$ theories with Lagrangian was suggested in \cite{Nazzal:2023wtw}\footnote{Our algorithm proposed in this paper relies on the perturbative expansion in RS parameters where 
corresponding $4d$ indices are well defined and have physical meaning. 
In this sense the spectrum we derived is strictly speaking valid only 
for the subspace of the full parameter space of RS model. In particular our results are blind to non-perturbative completion of the spectrum discussed in 
\cite{Hatsuda:2018lnv}.}. 
On the other hand, same integrable systems appear also in the study of \textit{ramified instanton partition functions} of $5d$ Kaluza-Klein (KK) theories \cite{Bullimore:2014upa,Bullimore:2014awa}. These are $5d$ instanton
partition functions with 1/2 BPS co-dimension two defects inserted. In particular, using the ramified instanton partition function a procedure to deduce the spectrum of
the RS model was conjectured in \cite{Hatsuda:2018lnv}. This procedure goes through construction of a function which depends on additional parameters,
specializing which to a discrete set of possibilities is conjectured to give 
the eigenfunctions. The choice of the specialization is often referred to as a 
choice of quantization, which generalizes the work in \cite{Nekrasov:2009rc}. 

In this note we will show that a particular choice of quantization, the so called B-model quantization introduced in \cite{Hatsuda:2018lnv}, leads to results consistent with the ones obtained from $4d$ index computations. We will perform explicit computations  for the two first eigenfunctions of the $A_1$ RS model and for the ground state of $A_2$ RS model.  We will also comment on how an alternative, A-model, quantization is related to the B-model quantization. From physics point 
of view the  results of this note should be useful for explicit computation of indices of theories which lack known Lagrangians in $4d$ by relating them to explicit instanton partition function computations of 5d KK theories. From the mathematical point of view it gives a concrete relation between two very 
different presentations of eigenfunctions of the $A_{N-1}$ RS elliptic integrable models. We expect that the results of this paper could be extended to address other integrable models appearing both in $4d$ and $5d$: {\it e.g.} 
the van Diejen model appearing in compactifications of the 6d E-string theory to 4d\cite{Nazzal:2018brc,Nazzal:2021tiu} and in the 
instanton claculations for $5d$ $USp(2N)$ gauge theory.
Unlike the RS model, in various generalizations very little is known about the spectrum and we expect that following through with generalizations
of our work should provide very concrete results in that direction.

This note is organized as follows. 
In  Section \ref{section:index} we  review the connection between 4d $\cN=1$ superconformal 
indices and elliptic integrable systems. We also summarize an algorithm for the derivation of the spectra of 
these operators from the index calculations that was proposed in
\cite{Nazzal:2023wtw}. Then in 
Section \ref{sec:instantons}
we  review the construction of ramified instanton partition functions from \cite{Bullimore:2014awa,Alday:2010vg,Kanno:2011fw}
and how they are related to the spectrum of RS model. In Section \ref{section:matching} we compare these two approaches 
and show that they give consistent results. In Section \ref{section:gluing} we discuss how different quantization choices 
in instanton calculations are related to the superconformal indices. Finally in Section \ref{section:discussion} we discuss 
possible future directions in which our results can be extended.

\section{Ruijsenaars-Schneider spectrum from the index.}
\label{section:index}

In this section we will very briefly review the main idea of \cite{Nazzal:2023wtw} on how to 
obtain ground state (and in some cases higher states as well)
of relativistic integrable models from 
the superconformal index calculations. 

Our starting point is the superconformal index of
$4d$ $\cN=1$ theory which is defined as the trace
over the Hilbert space of theory quantized on 
$S^3$:
\be
{\cal I}\left[T_{6d},{\cal C} \right]\left({\bf x},{\bf u}_{6d},p,q  \right)=
\mathrm{Tr}_{S^3} (-1)^F \, q^{j_2-j_1+\frac{R}2} \, p^{j_2+j_1+\frac{R}2}\, \prod_{\ell=1}^{\text{rank}\, G_F} x_\ell^{{\cal Q}_\ell}\,,
\label{index:def}
\ee
where $j_{1,2}$ are Cartan generators of $Spin(4)$, ${\cal Q}_\ell$ are charges of the global symmetry of $G_F$, $R$ is $R$-charge operator and $F$ is 
fermion number. Index depends on the fugacities $p,q,x_\ell$ that keep track of all the aforementioned charges of operators. Importantly in our index calculations we further assume 
\be
|p|<1\,,\quad |q|<1\,,
\label{pq:domain}
\ee
so that state counting problem is well defined and all expressions converge. 
From the point of view of our discussion of RS model this means that 
all our results obtained from indices are valid only in certain domain of parameters specified in \eqref{pq:domain}.

Four-dimensional theories we consider originate in the compactifications $6d$ SCFTs on a punctured Riemann surface. The resulting $4d$ theory depends 
on the original $6d$ SCFT, compactfication geometry which we denote as ${\cal C}$ and all the fugacities of $4d$ symmetries as denoted on the 
l.h.s. of \eqref{index:def}.  We have split the latter ones into fugacities $\bu_{6d}$ of global symmetries $G_{6d}$ of the original $6d$ SCFT 
and fugacities $\bx$ of $4d$ global symmetry emerging in the compactifications. We usually identify 
such symmetries $G_{5d}$  with the punctures of the 
Riemann surface ${\cal C}$. The subscript $5d$ is here since to obtain this symmetry one relies on an intermediate effective $5d$ gauge theory obtained 
in the circle compactification from $6d$ SCFT.

Given a particular $6d$ SCFT and punctures with certain $G_{5d}$ global symmetry one can \cite{Gaiotto:2012xa} obtain an 
elliptic relativistic  integrable model defined by a tower of commuting operators
\be
H_{\alpha}\left[T_{6d},G_{5d} \right]\left(\bx,\bu_{6d};p,q\right)\,.
\ee
To obtain this system one has to consider superconformal indices of $4d$ compactifications with insertion of surface defects. 
Index $\alpha$ of operators labels type of the defect inserted.
All operators act on the fugacities $\bx$ of the puncture symmetry $G_{5d}$ 
and depend on fugacities $\bu_{6d},p$ and $q$. 

An important property of these operators is that $\cN=1$ indices coming from the corresponding $T_{6d}$ compactifications are their \textit{Kernel functions}, i.e. 
\be\label{kernel}
&&H_\alpha\left[T_{6d},G_{5d}\right]({\bf x}_1,{\bf u}_{6d};q,p)\cdot {\cal I}[T_{6d}, {\cal C}](\{{\bf x}_1,{\bf x}_2,\cdots\},{\bf u}_{6d},q,p)=\\
&&\;\;\;\;\;\; \qquad\qquad H_\alpha\left[T_{6d},G_{5d}\right]({\bf x}_2,{\bf u}_{6d};q,p)\cdot {\cal I}[T_{6d},{\cal C}](\{{\bf x}_1,{\bf x}_2,\cdots\},{\bf u}_{6d},q,p)\,.\nonumber
\ee
Here $\bx_1$ and $\bx_2$ are fugacities of the two maximal punctures. These punctures can be of the same type, i.e. have the same $G_{5d}$, in which case operators on two sides
of equations are the same, or of different types, in which case operators are also different. 

Now we want to study the spectrum of operators $H_{\alpha}$: 
\be
H_\alpha\left[T_{6d},G_{5d}\right]({\bf x},{\bf u}_{6d};q,p)\cdot \psi_\lambda({\bf x}) =
E_{\alpha,\lambda}\, \psi_\lambda({\bf x})\,.
\label{operator:spectrum}
\ee
Index $\lambda$ of the eigenfunctions and  eigenvalues depends on a particular Hamiltonian. 
In some simplest cases, like $A_1$ RS model discussed in this paper,  it can be an integer. 
But more generally it should take the form of partition as it does in cases of Schur and Macdonald polynomials \cite{macdonald1998symmetric}. 
Since operators $H_{\alpha}$ are self-adjoint we can choose eigenfunctions $\psi_\lambda$ form orthonormal basis
w.r.t. certain measure
$\Delta\left(\bx,\bu_{6d};p,q\right)$
\be
\oint d{\bf x} \;\Delta({\bf x}, {\bf u}_{6d};q,p)\;\psi_\lambda({\bf x})\;\psi_{\lambda'}({\bf x}^{-1}) = \delta_{\lambda,\lambda'}\,.
\label{basis:ortho}
\ee
This measure  has a natural physical meaning as index contributions of certain multiplets required for gluing indices along punctures.
Operators $H_\alpha$ are  self-adjoint under this measure by construction which allows us to always choose orthonormal 
basis according to \eqref{basis:ortho}.

Due to the kernel property \eqref{kernel} it is natural to assume \cite{Gaiotto:2012xa} that the corresponding ${\cal N}=1$ indices have diagonal expansions in 
this basis. In particular index obtained in the compactification on the Riemann surface ${\cal C}$ with $s$ maximal punctures of the same type is
given by the following expansion:
\be
{\cal I}[T_{6d},{\cal C}](\{{\bf x}_j\},{\bf u}_{6d},q,p)=
\sum_{\lambda\in \Lambda} C_\lambda[T_{6d},{\cal C}]({\bf u}_{6d};q,p)
\prod_{j=1}^s\,\psi_{\lambda}({\bf x}_j)\,,
\label{index:expansion}
\ee
where $\Lambda$ denote the set of all possible $\lambda$ indexing eigenfunctions. In order to make sense of this expression we 
assume that there is a natural ordering of $\lambda \in \Lambda$ so we can enumerate them in such a way that: 
\be
\lambda_0 \leq \lambda_1 \leq \lambda_2 \leq ...\,. 
\ee
The idea of \cite{Nazzal:2023wtw} is to derive the eigenfunctions $\psi_{\lambda_i}$ of elliptic operators as expansion in 
$p$ and $q$ fugacities using simple facts summarized above. The starting point of the construction is index of the theory 
obtained in compactification on the surface with at least two maximal punctures and some flux ${\cal F}$. Additionally a Riemann surface can have some 
genus and a number of other punctures. This index according to what is said above can be written in the following form 
\be
{\cal I}_1({\bf x}_1,{\bf x}_2) = \sum_{i=0}^\infty C_{\lambda_i}\; \psi_{\lambda_i}({\bf x}_1)\; \psi_{\lambda_i}({\bf x}_2)\,.
\label{tube:index:general}
\ee
If we glue $n$ copies of such theories along maximal punctures we obtain compactification on a surface still with two maximal punctures 
but with flux, genus and number of other punctures all multiplied by $n$. Performing gluing and using orthonormality condition \eqref{basis:ortho} 
we can derive the corresponding index given by: 
\be
{\cal I}_{n} ({\bf x}_1,{\bf x}_2) = \sum_{i=0}^\infty \left(C_{\lambda_i}\right)^{n}\; \psi_{\lambda_i}({\bf x}_1)\; \psi_{\lambda_i}({\bf x}_2)\,.
\ee
Now we make an important assumption that lowest order of $C_{\lambda_i}$ in $p$ and $q$ expansion grows with $i$. This assumption 
will be justified by self-consistency of our calculations. Then up to any set order in $p,q$ expansion  only the ground state will contribute 
to the index starting from some value of $n$. In particular in the limit $n\to \infty$ we can compute the following quantity:
\be \label{Eq:C0}
C_{0}\equiv C_{\lambda_0}= \lim_{n\to \infty} \frac{{\cal I}_{n+1} ({\bf x}_1,{\bf x}_2)}{{\cal I}_{n} ({\bf x}_1,{\bf x}_2)}\,.
\ee
And from here evaluate the ground state eigenfunction
\be
\label{Eq:psi0}
\widetilde \psi_0({\bf x})\equiv \psi_0({\bf 1})\; \psi_0({\bf x})= \lim_{n\to \infty}\frac{1}{\left(C_0\right)^{n}}\,
{\cal I}_{n} ({\bf x},1)\,.
\ee
This argument can be extended to all states in the spectrum. For example for the first excited state we get: 
\be\label{Eq:C1}
&&C_{1}\equiv C_{\lambda_1}= \lim_{n\to \infty} \frac{{\cal I}_{n+1}(x,1)-(C_0)^{n+1}
\widetilde \psi_0(x)} {{\cal I}_{n} (x,1)-(C_0)^{n}\widetilde \psi_0(x)}\,.\\
&&\label{Eq:psi1}
\widetilde \psi_1({\bf x})\equiv \psi_1({\bf 1})\; \psi_1({\bf x})= \lim_{n\to \infty}\frac{1}{\left(C_1\right)^{n}}\,
\left({\cal I}_{n} ({\bf x},1)-(C_0)^{n}\widetilde \psi_0(x)\right)\,.
\ee
Finally it is worth mentioning that eigenfunctions $\psi_i(\bf{x})$ are by construction orthonormal under the measure $\Delta\left(\bf{x},\bf{u}_{6d};q,p\right)$. 
Since this normalization is the natural one and simplifies some calculations we will further work with $\psi_i(\bf{x})$ function.
To obtain it back from the $\widetilde \psi_i(\bf{x})$ 
we need to normalize it as follows 
\be
\psi_i({\bf x})=\frac{\widetilde\psi_i({\bf x})}{\sqrt{\widetilde\psi_i({\bf 1})}}
\ee

The construction presented above is general and can be applied in many different compactification settings \cite{Nazzal:2023wtw}. 
In present paper we will concentrate on two examples where calculations can be done both using our method and 
ramified instanton partition functions discussed later in Section \ref{sec:instantons}. These are $A_1$ and $A_2$ RS models. 
Unfortunately due to the calculation complexity it
is very hard to perform similar calculations for the 
higher rank RS models but we believe that the general conclusions 
of our paper should also hold for all $A_N$ RS models.

\subsection{$A_1$ Ruijsenaars-Schneider model} 
\label{sec:A1RS}

We start our considerations with the simplest example of elliptic integrable models, $A_1$ Ruijsenaars-Schneider system.
This model is defined by a single Hamiltonian whose action on a trial function is given by:
\be
{\cO}\cdot \psi(x)=\frac{\thf{\sqrt{\frac{p}{q}}tx^{-2} }}{\thf{x^2}}\psi\left(q^{1/2}x \right)+\frac{\thf{\sqrt{\frac{p}{q}}tx^{2} }}{\thf{x^{-2}}}\psi\left(q^{-1/2}x \right)\,,
\label{RS:hamiltonian:1}
\ee
which is slightly different from the canonical parametrization usually used in literature. In the setting described above this model emerges when we study 
superconformal indices of class $\mathcal{S}$ theories obtained in the compactificaitons of $6d$ $(2,0)$ SCFT with $G_{6d}=\SU(2)$ global symmetry. Parameter $t$ in 
\eqref{RS:hamiltonian:1} plays the role of the fugacity for this $6d$ global symmetry. Parameters $p$ and $q$ as usually stand for the fugacities of the Cartan of $Spin(4)$. 

The crucial element we need for our construction is an integration measure $\Delta\left(\bf{x},\bf{u}_{6d};q,p\right)$ under which RS Hamiltonian 
\eqref{RS:hamiltonian:1} is self-adjoint as discussed in the previous Section. In particular this measure takes the following simple form 
\be
\Delta\left(x,t;q,p \right)=\frac{\qPoc{q}{q}\qPoc{p}{p}}{2}\frac{\GF{\sqrt{pq}\,t^{-1}x^{\pm 2}}\GF{\sqrt{pq}\,t^{-1}}}{\GF{x^{\pm 2} }}\,.
\label{A1:index:measure}
\ee
One more thing we would like to make before proceeding is to reduce our operator \eqref{RS:hamiltonian:1} to the canonical form of 
the $A_1$ Ruijsenaars-Schneider model 
\be
{\cal H}_{RS}\cdot \psi(x)=\frac{\thf{\pq{1}{2}tx^{2} }}{\thf{x^2}}\cdot\psi\left(q^{1/2}x \right)+
	\frac{\thf{\pq{1}{2}tx^{-2} }}{\thf{x^{-2}}}\cdot\psi\left(q^{-1/2}x \right)\,,
\label{RS:hamiltonian:2}
\ee
 It can be easily seen that the two operators are related by simple conjugation
\be
\cO=\KK^{-1}\cdot {\cal H}_{RS}\cdot \KK
\label{operator:conjugation}
\ee
where the factor $\KK$ is given by 
\be
\KK=\GF{\pq{1}{2}tx^{\pm 2}}^{-1}
\label{K:factor}
\ee
Hence the eigenfunctions $\psi_i(x)$ of \eqref{RS:hamiltonian:1} that we derive in our index computations are related to the
eigenfunctions $\psi^{RS}_i(x)$ of the canonical RS model \eqref{RS:hamiltonian:2} as follows:
\be
\psi_i^{RS}(x)=\KK\cdot\psi_i(x)\,.
\label{RS:eigenfunctions:conj}
\ee

Using expressions \eqref{Eq:psi0} and \eqref{Eq:psi1} we obtain the following eigenfunctions for the first two energy levels: 

\be
\psi_0^{RS}(x)&=&1+\sqrt{pq}\left(t+t^{-1}\left(x^2+x^{-2}\right)\right)+pq \left[ \frac{1}{2}\left(t^2+t^{-2}\right)-2+t^{-2}\left(x^4+x^{-4} \right)\right]+
\nn\\
&&\hspace{30mm}\sqrt{pq}(p+q)\left[t^{-1}\left(x^2+x^{-2} \right)+\frac{1}{2}t^{-1}+\frac{3}{2}t+\dots \right]\,,
\nn\\
\psi_1^{RS}(x)&=&\left(x+\frac{1}{x}\right)\left[1+\sqrt{pq}\left(\frac{1}{2}t-\frac{1}{2}t^{-1}+t^{-1}\left( a^2+a^{-2} \right)\right)+\frac{p}{2}+\frac{q}{2}  + 
\frac{3}{8}\left(p^2+q^2\right)-
\right.\nn\\
&&\left.
pq\left(1+\frac{1}{2}\left(x^2+x^{-2} \right)-\frac{7}{8}t^{-2}-\frac{3}{8}t^2-t^{-2}\left(x^4+x^{-4}\right)+\frac{t^{-2}}{2}\left(x^2+x^{-2}\right)\right)+
\right.\nn\\
&&\left.
 \hspace{3cm}	\sqrt{pq}(p+q)\left(\frac{5}{4}t-\frac{t^{-1}}{4}+\frac{3}{2}t^{-1}\left(x^2+x^{-2}\right)\right)+\dots
\right]\,.
\label{A1RS:eigfunct:result}
\ee
Expansions above are performed in parameter
$(pq)^{1/4}$ with $p/q$ ratio fixed. Dots 
correspond to higher orders in this expansion.

Substituting eigenfunctions \eqref{A1RS:eigfunct:result} into eigenvalue equation \eqref{operator:spectrum} with the $A_1$ RS Hamiltonian \eqref{RS:hamiltonian:2} we can obtain the 
ground state eigenvalue given by: 
\be
E_0=1-p+\left(t+\frac{1}{t}\right)\sqrt{pq}-pq+\left( t+\frac{1}{t}\right)p\sqrt{pq}-p^2+\dots\,.
\label{A1RS:ground:state:energy}
\ee

An important check of our results is the Macdonald limit of the problem. This limit in our notation corresponds 
to first rescaling $t$ fugacity, $t\to t/\sqrt{pq}$ and then taking $p\to 0$ limit. In this case RS Hamiltonian \eqref{RS:hamiltonian:2} reduces 
to the Macdonald operator whose spectrum is known exactly. In particular its eigenfunctions have simple polynomial form and are known as 
\textit{Macdonald polynomials}\cite{macdonald1998symmetric}. This fact has played an important role in index computations in the past \cite{Gadde:2011uv,Gadde:2009kb,Gadde:2011ik}.
In Appendix \ref{app:A1:Macdonald} we summarize an explicit form of $A_1$ Macdonald polynomials as well as eigenvalues of the 
Macdonald operator. 

We would also like to fix the relation between our eigenfunctions \eqref{A1RS:eigfunct:result} in the Macdonald limit and normalized $A_1$ Macdonald polynomial
\eqref{rogers:polynom:norm}. Functional dependence of these two obviously should be the same since they are both eigenfunctions of the Macdonald 
operator \eqref{A1:Macdonald:operator}. However their overall normalization differs since in Macdonald limit our functions are orthonormal w.r.t. the 
following integration measure: 
\be
\widetilde{\Delta}(x,t;q)\equiv \lim_{p\to 0}\left[\GF{tx^{\pm 2}}^2\Delta\left(x,t\left(pq\right)^{-1/2};q\right)\right]=
\frac{1}{2}\qPoc{q}{q}\qPoc{t}{q}\frac{\qPoc{x^{\pm 2}}{q}}{\qPoc{tx^{\pm 2}}{q}}\,,
\ee
where we start with the integration measure \eqref{A1:index:measure} add conjugation factor \eqref{K:factor} and finally perform the Macdonald limit 
as described above. 
As we see measure we got in the end has an extra constant factor w.r.t. the canonical Macdonald polynomial measure as specified in \eqref{A1:macdonald:measure}. 
This difference can be fixed by rescaling the eigenfunctions themselves. Hence we can relate the Macdonald limit of our wavefunctions \eqref{A1RS:eigfunct:result}
with the normalized Macdonald polynomials \eqref{rogers:polynom:norm} in the following way:
\be
\psi_n^{RS}(x)=\frac{P_n(x;t,q)}{\sqrt{\qPoc{q}{q}\qPoc{t}{q}}}
\label{A1:RS:index:relation}
\ee
Using explicit expressions for the eigenfunctions \eqref{A1RS:eigfunct:result} and eigenvalue \eqref{A1RS:ground:state:energy} and performing 
Macdonald limit we can check that they indeed reproduce results \eqref{A1:macdonald:levels12} and \eqref{ground:state:macdonald} 
providing us a crosscheck of our results.


\subsection{$A_2$ Ruijsenaars-Schneider model} 
\label{sec:A2RS}

The results for the $A_1$ RS model summarized in the previous section has already been presented in our paper \cite{Nazzal:2023wtw}.
Here we extend these results to the higher rank  $A_2$ RS model. The model is defined by the tower of Hamiltonians with the action given by: 
\be
{\cO}_{A_1}^{(r,s)}\cdot \psi(x_i) &=& \sum\limits_{\substack{n_1+n_2\\+n_3 = r}}\sum\limits_{\substack{m_1+m_2\\+m_3 = s}}\psi\left(q^{\frac{s}{3}-m_i}p^{\frac{r}{3}-n_i}x_i\right)
\prod\limits_{i,j=1}^3\left[\prod\limits_{m=0}^{m_i-1}\frac{\thf{p^{n_j+\frac{1}{2}}q^{m+m_j-m_i+\frac{1}{2}}t\frac{x_i}{x_j}}}
{\thf{p^{-n_j}q^{m-m_j}t\frac{x_j}{x_i}} } \right]\times
\nn\\
&& \hspace{4cm} \left[\prod\limits_{n=0}^{n_i-1}\frac{\thfq{p^{n+n_j-n_i+\frac{1}{2}}q^{m_j-m_i+\frac{1}{2}}t\frac{x_i}{x_j}}}
{\thfq{p^{n-n_j}q^{m_i-m_j}t\frac{x_j}{x_i}} } \right]\,.
\label{A2:RS:action}
\ee
Of these only first two operators $(1,0)$ and $(2,0)$ are truly independent. Here we use the same non-standard choice of parameters as in 
$A_1$ case discussed in the previous section. Integration measure we use in this case is given by 
\be
\Delta(x_i,t;p,q)= \frac{\qPoc{p}{p}^2\qPoc{q}{q}^2}{3!}\GF{\sqrt{pq}t^{-1}}^2\prod\limits_{i\neq j}^3\frac{\GF{\sqrt{pq}t^{-1}\frac{x_i}{x_j}}}
{\GF{\frac{x_i}{x_j}}}
\label{A2:RS:measure}
\ee
Finally once again in order to transform our Hamiltonian \eqref{A2:RS:action} to the standard form of RS we should conjugate it 
in a following way: 
\be
{\cal H}_{RS}=\KK_{A_2}\cdot\cO\cdot \KK_{A_2}^{-1}
\label{operator:conjugation:A2}
\ee
where the conjugation factor $\KK_{A_2}$ is given by 
\be
\KK_{A_2}=\prod\limits_{i\neq j}^3\GF{\pq{1}{2}t\frac{x_i}{x_j}}^{-1}
\label{K:factor:A2}
\ee
Then given an eigenfunction of our operator \eqref{A2:RS:action} corresponding eigenfunction of $A_2$ RS Hamiltonian 
would be given by:
\be
\psi_{\lambda}^{RS}(x) = \KK_{A_2}\cdot \psi_{\lambda}(x)\,,
\label{A2:RS:eigfunct:rel}
\ee
where eigenfunctions are labelled by partitions $\lambda$. 

As usually in order to derive eigenfunctions of \eqref{A2:RS:action} we should start with the index of the tube theory \eqref{tube:index:general}. In case of $A_2$ RS Hamiltonian it takes the following form \cite{Gaiotto:2012xa}: 
\be
\cI_2(x,y)=\prod\limits_{i,j=1}^3\GF{\pq{1}{4}\sqrt{t}\left(x_iy_j  \right)^{\pm 1} }\,,
\label{A2:tube:index}
\ee
which is just the index of free bifundamental hypermultiplet. 
Using this tube index in our algorithm of finding spectrum of finite difference operators 
we arrive to the following result for the ground state wavefunction corresponding to an empty partition:
\be
\psi_{\emptyset}^{RS}(x_1,x_2,x_3) = 1 + \sqrt{pq}\left[2t + \frac{1}{t}\left(\frac{x_1}{x_2}+\frac{x_2}{x_1}+\frac{x_3}{x_2}+\frac{x_2}{x_3}+
\frac{x_3}{x_1}+\frac{x_1}{x_3} \right) \right]+
\nn\\
\sqrt{pq}(p+q)\left[\frac{5}{2}t+\frac{1}{2}t^{-1}+\frac{1}{t}\left(\frac{x_1}{x_2}+\frac{x_2}{x_1}+\frac{x_3}{x_2}+\frac{x_2}{x_3}+
\frac{x_3}{x_1}+\frac{x_1}{x_3} \right)   \right]+ \dots
\label{A2:RS:ground:state}
\ee
Here we show just a few first orders in the expansion performed in 
$(pq)^{1/4}$ parameter with $p/q$ ratio fixed.
The full result we obtained is valid up
to $(pq)^{11/4}$ order. Unfortunately this is not 
large enough neither to obtain the ground state energy of the $A_2$ RS Hamiltonian \eqref{A2:RS:action} nor to go higher in the spectrum and 
derive first excited eigenfunction as we did in $A_1$ case.

\section{Ramified Instantons.}
\label{sec:instantons}

Another way to derive the spectrum of RS operators is using \textit{ramified instanton partition function}
as was proposed in \cite{Bullimore:2014awa,Hatsuda:2018lnv}
\footnote{Later some generaliaztions of this construction to other integrable system was considered in 
\cite{Koroteev:2018isw,Koroteev:2019gqi,Gorsky:2021wio}}. 
These are just instanton partition functions of $5d$ $\U(N)$ or $\SU(N)$ $\NN=1^*$ SYM theory
on the $\Omega$ background $S^1\times\mathbb{R}^4_{\epsilon_1,\epsilon_2}$ with the insertion of the 1/2 BPS monodromy defects 
of Gukov-Witten type \cite{Gukov:2006jk}. 
Each monodromy defect is labeled by the partition $\rho=[n_1,n_2,\dots,n_s]$, where $\sum_{i=1}^sn_i=N$ and without loss of generality 
we choose $n_i$ to be ordered $n_1\geq n_2\geq \dots \geq n_s$. 
The full gauge group in this case is broken down to $S[\U(n_1)\times\U(n_2)\times\dots\times\U(n_s)]$.
Without the defect topological sectors of the theory are labelled by a single instanton number $k$. 
Once we introduce the defect we should also introduce instanton number for each of $s$ subgroups our original gauge group was broken to.
Hence we label different sectors with the set of $s$ integers 
$\{k_1,\dots,k_s\}$ that should add up to $k=\sum_{j=1}^sk_j$. To obtain the full instanton
partition function in the presence of the defect, we should sum over sectors as follows:
\be
Z_{\rho}=\sum\limits_{\vl}Q_1^{k_1(\vl)}\cdot\dots\cdot Q_s^{k_s(\vl)}Z_{\rho,\vl}\,.
\label{5d:inst:ramified}
\ee
Here each sector is labelled by Young tableaux
\be
\vl = \{\lambda_{j,\alpha}\}\,,\quad j = 1,\dots,s\,,\quad \alpha = 1,\dots,n_s\,,
\ee
where the boxes in the $i$-th column of $\lambda_{j,\alpha}$ contribute to the instanton number $k_{i+j-1}$ for $i+j-1$ mod $N$. 
Instanton parameters $Q_j$ of different sectors are defined so that 
\be
Q_1\cdot\dots\cdot Q_s=Q\,.
\ee
The contribution $Z_{\rho,\vl}$ for each topological sector, corresponding to a given Young tableau, can be calculated using a simple orbifolding procedure applied to the standard instanton ADHM data. This approach has been explored in \cite{Alday:2010vg,Kanno:2011fw}. Specific expressions for this contribution, $Z_{\rho,\vl}$, as well as detailed derivations of the ramified instanton 
partition functions are available in Section 4.2 of \cite{Bullimore:2014awa}. Our notations in this section follows those of the same reference. 

Parameters which ramified instanton partition functions depend on are
\be
\mu_j\equiv e^{2\pi i a_j}\,,\qquad q\equiv e^{2\pi i\epsilon_1}\,,\qquad \eta^2=qe^{-2\pi im}\,,\quad j=1,\dots,N\,.
\label{instanton:exp:parameters}
\ee
where $\epsilon_{1,2}$ are equivariant parameters of $\Omega$ deformation, $a_i$ are Coulomb branch parameters and $m$ is the mass of the adjoint hypermultiplet.

\subsection{$A_1$ RS from instantons.}

Now we concentrate on the simplest example of $\SU(2)$ theory.
In this case it is convenient to introduce
the following notations for the instanton parameters of two sectors:
\be
(+)\quad Q_1=z\,,\qquad Q_2=\frac{Q}{z}\,,
\nn\\
(-)\quad Q_1=\frac{Q}{z}\,,\qquad Q_2=z\,.
\label{a1:instanton:parameters}
\ee
These two sectors correspond to two supersymmetric vacua in the 3d theory on the defect, specifically the 3d $\mathcal{N}=4$ $T[SU(2)]$ theory, when generic parameters $a_j, m$, as well as a FI parameter $z$, are turned on.

We will be particularly interested in the Nekrasov-Shatashvili (NS) limit $\epsilon_2\to 0$. The ramified instanton partition function is divergent in this limit, so
following \cite{Bullimore:2014awa} we use the normalized expectation value of the monodromy defect
\be
D_{[1,1]}^{(\pm)}\equiv\lim_{\epsilon_2\to 0}\frac{Z_{[1,1]}^{(\pm)}}{Z}\,,
\label{monodromy:expactation}
\ee
where $Z$ is the instanton partition function without defect and $Z_{[1,1]}^{(\pm)}$ is the ramified instanton partition function as defined in 
\eqref{5d:inst:ramified} for the partition $\rho = [1,1]$. Superscripts $(\pm)$ correspond to the two SUSY vacua associated to instanton 
parameters in \eqref{a1:instanton:parameters}.
Using the specific expressions provided in Section 4.2 of \cite{Bullimore:2014awa}, one can expand the expectation values of the monodromy defect in terms of $z$ and $Q/z$. The first few
orders in the expansion are given by
\be
D_{[1,1]}^{(+)}&=&1+\frac{q(\eta^2-1)(\eta^2 \mu_2-\mu_1)}{\eta^2(q-1)(\mu_2q-\mu_1)}z+\frac{q(\eta^2-1)(\eta^2 \mu_1-\mu_2)}{\eta^2(q-1)(\mu_1q-\mu_2)}\frac{Q}{z}+\dots\,,
\nn\\
D_{[1,1]}^{(-)}&=&\left.D_{[1,1]}^{(+)}\right|_{\mu_1\leftrightarrow \mu_2 }\,,
\label{ramified:inst:u2}
\ee
where we have used exponentiated parameters \eqref{instanton:exp:parameters}.
As was noticed in \cite{Bullimore:2014awa} these expressions satisfy
the following finite difference equations
\be
\left[\mu_1\frac{\thfQ{\eta^2\frac{\tau_1}{\tau_2}}}{\thfQ{\frac{\tau_1}{\tau_2}}}p_{\tau_1}
+\eta^2\mu_2 \frac{\thfQ{\eta^2\frac{\tau_2}{\tau_1}}}{\thfQ{\frac{\tau_2}{\tau_1}}}p_{\tau_2}\right]D^{(+)}_{[1,1]}=E_{(1)}D^{(+)}_{[1,1]}\,,
\nn\\
\left[\mu_2\frac{\thfQ{\eta^2\frac{\tau_1}{\tau_2}}}{\thfQ{\frac{\tau_1}{\tau_2}}}p_{\tau_1}
+\eta^2\mu_1 \frac{\thfQ{\eta^2\frac{\tau_2}{\tau_1}}}{\thfQ{\frac{\tau_2}{\tau_1}}}p_{\tau_2}\right]D^{(-)}_{[1,1]}=E_{(1)}D^{(-)}_{[1,1]}\,,
\label{A1:inst:RS:action1}
\ee
where $p_{\tau_i}$ are shift operators $\tau_i\to q\tau_i$ and we introduced $\tau_{1,2}$ parameters using $z=\tau_2/\tau_1$.

The reason why the defect partition functions satisfy these difference equations is as follows. The 3d $\mathcal{N}=4$ theory on the monodromy defect has a set of supersymmetric vacua which, when coupled to a 5d theory, are non-trivially fibered over the Coulomb branch of the bulk 5d gauge theory. This implies that the twisted chiral ring relations, which define the massive vacua, play the same role as the Seiberg-Witten curve for the bulk theory, as suggested in \cite{Gaiotto:2013sma}. Moreover, when the $\Omega$-parameter $\epsilon_1$ is turned on, it leads to the quantization of the twisted chiral ring relations, with $\epsilon_1$ acting as the Planck constant. The above difference equations are precisely these quantized versions of the twisted chiral ring relations, as suggested in \cite{Bullimore:2014awa}.

The eigenvalues $E_{(1)}$ in the equations above are given by an expectation value $\langle W_{(1)}^{\U(N)}\rangle$ of the supersymmetric Wilson loops 
in the fundamental representation wrapping $S^1$ cycle. 
For $\SU(N)$ gauge groups we should also divide by a $\U(1)$ Wilson loop expectation value: 
\be
\langle W_{(1)}^{\SU(N)}\rangle=\frac{\langle W_{(1)}^{\U(N)}\rangle}{\langle W_{(1)}^{\U(1)}\rangle}\,,
\label{sun:wils:loop}
\ee
and impose $\SU(N)$ condition $\prod\limits_{i=1}^N\mu_i=1$ on the Coulomb branch parameters. The $U(1)$ contribution is given by 
\be
\langle W_{(1)}^{\U(1)}\rangle=\frac{\left( Q/\eta^{2};Q \right)_{\infty}\left(\eta^2 Q/q;Q \right)_{\infty} }{\left( Q;Q \right)_{\infty}\left(Q/q;Q \right)_{\infty}}
\label{u1:wilson:loop}
\ee
These Wilson loop expectation values can be computed by using the instanton partition function with a universal bundle inserted due to the Wilson loop operator \cite{Bullimore:2014awa}, or by applying localization to the 1d ADHM quantum mechanics living on the worldline of the Wilson loop operator  \cite{Kim:2016qqs}. In the simplest case of the $\U(2)$ gauge theory the result of these calculations is given by:
\be
E_{(1)}^{\U(2)}&\equiv &\langle W_{(1)}^{\U(2)}\rangle=(\mu_1+\mu_2)\left[1-
	\right.\nn\\
	&&\hspace{-7mm} \left.
	(1-\eta^2)(q-\eta^2)\frac{\mu_1\mu_2\left( \eta^2+q\left( \eta^4+\eta^2+q \right)\right)-
	(\mu_1+\mu_2)^2\eta^2q}{\eta^4q(\mu_1q-\mu_2)(\mu_2q-\mu_1)}Q+O(Q^2)\right] \ .
\label{u2:wilson:loop}
\ee
It can be directly checked by substitution into \eqref{A1:inst:RS:action1} that this is indeed an eigenvalue of $A_1$ RS 
Hamiltonian.

We can notice now that equations \eqref{A1:inst:RS:action1} are not exactly of the form of RS Hamiltonian action \eqref{RS:hamiltonian:2} 
due to prefactors invloving Coulomb branch parameters $\mu_{1,2}$ and hypermultiplet mass $\eta$. In order to obtain canonical 
$A_1$ RS action we need to add to the ramified instanton partition functions $D_{[1,1]}^{(\pm)}$ prefactors of the following form:
\be
D_{[1,1]}^{(\pm)}(z) =z^{-\frac{1}{2}\log_q\left[\eta^2\left(\frac{\mu_2}{\mu_1}\right)^{\pm 1 }\right] }
D^{(\pm)}\left(z\right)
\label{a1:ramified:instanton:prefactor}
\ee
the resulting functions become eigenfunctions of the canonical $A_1$ RS Hamiltonian: 
\be
\left[\frac{\thfQ{\eta^2 z}}{\thfQ{z}}p_{z}+
\frac{\thfQ{\eta^2 z^{-1} }}{\thfQ{z^{-1}}}p_{z}^{-1} \right]D^{(\pm)}= \frac{E_{(1)}}{\eta\sqrt{\mu_1\mu_2}}D^{(\pm)}\,,
\label{u2:rs:instanton}
\ee
where we have also made a substitution $z=\tau_2/\tau_1$. In this form the Hamiltonian looks almost identical to the Hamiltonian 
\eqref{RS:hamiltonian:2}  considered in the previous section. Notice that in equation above we only obtain a \textit{formal eigenfunction} of 
$A_1$ RS Hamiltonian. In particular the function $D^{(\pm)}$ in 
\eqref{u2:rs:instanton} is an eigenfunction of $A_1$ RS Hamiltonian. However it is not $L^2$ normalizable on $\mathbb{R}$ in general. Hence 
it can not be in the spectrum of RS. Nevertheless ramified instanton partition functions, and hence also $D^{(\pm)}$, depend on the Coulomb
branch parameters which RS Hamiltonian does not depend on. Hence we are free to choose them as we want. 
In order to fix normalizability problem it is natural then to try find such values of Coulomb branch parameters that the 
eigenfunctions we obtain are normalizable. In other words we need to find proper \textit{quantization conditions} 
for the Coulomb branch parameters $\mu_i$. 

A natural quantization condition for the Coulomb branch parameters in 5d $\mathcal{N}=1^*$ theories was proposed in \cite{Crichigno:2018adf}. The main idea here is to place the 5d theory on a compact space $S^3\times S^2$, and evaluate the partition function using the saddle point method in the limit $\epsilon_2\rightarrow0$, which implements the A-twist along $S^2$, where $\epsilon_2$ is the background field for the $SU(2)$ isometry on $S^2$. The saddle point analysis yields Bethe Ansatz equations written in terms of the Coulomb branch parameters $\mu_i$, whose solutions quantize these parameters to a discrete set of values. If we naively apply the quantization condition proposed in \cite{Nekrasov:2009rc} directly to relativistic Hamiltonian systems, the associated defect partition function would exhibit infinitely many poles in the Planck constant $\hbar$ (or the $\Omega$ parameter $\epsilon_1$), showing an inconsistency. However, this issue is naturally resolved when employing the quantization condition derived from the partition function on $S_b^3\times S^2$. Here, the Planck constant is identified with the $S^3$ squashing parameter $\hbar=b$, and such poles of $\hbar$ are no longer present when the theory is formulated on a compact manifold like $S^3\times S^2$. 

Practically, under equivariant localization, the partition function on $S^3\times S^2$ is factorized into two contributions, one expressed with $b$ and the other one with $b^{-1}$, from the north and south poles of $S^2\subset S^3$. This combination of contributions effectively accomplishes the non-perturbative completion in $\hbar$ of the exact quantization condition for the elliptic RS model, specifically the \textit{A-model quantization condition}, proposed in \cite{Hatsuda:2018lnv}. However, solving the A-model quantization condition is highly challenging in practice because the Bethe Ansatz equations involve an infinite sum of instanton contributions. In particular, the expansion using A-model parameters takes place in a different parameter regime than that used for the 4d index expansion, and thus this requires a non-perturbative resummation process to derive the 4d index expression, which is a daunting task.

Due to challenges associated with the A-model approach, which relies on solving the Bethe Ansatz equations from the saddle point method, we will instead adopt the \textit{B-model quantization condition} proposed in \cite{Hatsuda:2018lnv}. There are two primary differences between the A-model and B-model quantizations. First, the B-model quantization condition for the Coulomb branch parameters is notably simpler. In $\SU(2)$ case it takes the following form:
\be
\mu_2=\frac{1}{\mu_1} = \eta^{\mp 1}q^{\mp \frac{n}{2}}\,,\quad n \in \mathbb{Z}_{\geq 0}
\label{B-model:quantization}
\ee
where different signs $\mp$ should be used upon quantizing Coulomb branch parameters inside $D^{(+)}$ or $D^{(-)}$ partition functions 
correspondingly. This choice of signs doesn't affect the value of the eigenvalue $E_{(1)}$  since it is symmetric with respect to 
$\mu_1 \leftrightarrow \mu_2$ exchange, which is the $SU(2)$ Weyl reflection. Notice that this quantization condition \eqref{B-model:quantization} does not depend on the instanton parameter $Q$ meaning 
it is not sensitive to the non-perturbative corrections.
Second, the partition function with or without a monodromy defect in the B-model quantization is represented by a single factor of the partition function on $S^1\times \mathbb{R}^4$. In contrast, the A-model quantization involves two factors coming from the north and south poles of $S^2$ within $S^3$. The equivalence of these two quantization conditions, once the parameters are appropriately identified, was numerically verified in \cite{Hatsuda:2018lnv}. In fact, these two quantization conditions are related with each other through an S-duality transformation, which we will discuss further in Section \ref{section:gluing} in more details.

Last thing we would like to discuss here is the Macdonald limit of ramified instanton partition functions. We start with the normalized expectation value $D_{[1,1]}^{(\pm)}$ for the monodromy defect as defined in 
\eqref{monodromy:expactation}. In terms of instanton calculations 
Macdonald limit would correspond to taking instanton counting number to 
zero $Q\to 0$. This can be easily seen from the form of RS Hamiltonian 
\eqref{u2:rs:instanton} since it reduces to the Macdonald operator 
\eqref{A1:Macdonald:operator} in $Q\to 0$ limit. If we now consider monodromy 
expectation value in this limit we kill all instanton corrections 
and end up with the vortex partition function of $3d$ $T\left[\SU(2)\right]$ 
theory living on the defect. They are given by $q$-hypergeometric series:
\be
P_{[1,1]}^{(+)}\equiv\lim_{Q\to 0}D_{[1,1]}^{(+)}= ~_2F_1\left(\eta^2,\,\eta^2\frac{\mu_2}{\mu_1},\,q\frac{\mu_2}{\mu_1},\,q,\,q\eta^{-2}z\right)\,,
\nn\\
P_{[1,1]}^{(-)}\equiv\lim_{Q\to 0}D_{[1,1]}^{(-)}= ~_2F_1\left(\eta^2,\,\eta^2\frac{\mu_1}{\mu_2},\,q\frac{\mu_1}{\mu_2},\,q,\,q\eta^{-2}z\right)\,,
\ee
where $q$-hypergeometric series are defined as follows: 
\be
_2F_1(a,b,c,q,z)=\sum\limits_{k=1}^{\infty}\frac{\qPocn{a}{q}{k}\qPocn{b}{q}{k}}{\qPocn{c}{q}{k}\qPocn{q}{q}{k}}z^k\,.
\label{q:hypergeometric:def}
\ee
Without loss of generality let's now concentrate on $D^{(-)}_{[1,1]}$ 
function. All derivations for another, $(+)$ choice of vacua work in a similar way. 
This function as specified above has representation of an infinite series which is quite different from the 
expected Macdonald polynomial given in \eqref{rogers:polynom} and \eqref{rogers:polynom:norm}. 
In order to obtain finite polynomial we need to impose B-model quantization conditions \eqref{B-model:quantization}. 
In this case it can be shown that $q$-hypergeometric series truncate at $k=n$  and become finite polynomial: 
\be
\left.P_{[1,1]}^{(-)}\right|_{\frac{\mu_2}{\mu_1}=\eta^2q^n}=\frac{\qPocn{q}{q}{n}}{\qPocn{\eta^2}{q}{n}}
\sum\limits_{k=1}^{n}\frac{\qPocn{\eta^2}{q}{k}\qPocn{\eta^2}{q}{n-k}}{\qPocn{q}{q}{k}\qPocn{q}{q}{n-k}}z^k
\ee
This is almost an expression we need. Final step in order to obtain Macdonald polynomials is to include the 
prefactor in \eqref{a1:ramified:instanton:prefactor}. Notice that this prefactor does not change when we take 
$Q\to 0$ limit. However when we impose B-model quantization condition \eqref{B-model:quantization} it simplifies down to an integer power of $z$ so 
we obtain: 
\be
&&P^{(-)} \equiv z^{\frac{1}{2}\log_q\left[\eta^2\frac{\mu_1}{\mu_2}\right] }P_{[1,1]}^{(-)}\,,
\nn\\
&&\left.P^{(-)}\right|_{\frac{\mu_2}{\mu_1} = \eta^2q^n}=\frac{\qPocn{q}{q}{n}}{\qPocn{\eta^2}{q}{n}}
\sum\limits_{k=1}^{n}\frac{\qPocn{\eta^2}{q}{k}\qPocn{\eta^2}{q}{n-k}}{\qPocn{q}{q}{k}\qPocn{q}{q}{n-k}}z^{k-\frac{n}{2}}=
\frac{\qPocn{q}{q}{n}}{\qPocn{\eta^2}{q}{n}}R_n\left(\sqrt{z};\eta^2,q\right)\,,
\ee
where $R_n(x;t,q)$ are  non-normalized $A_1$ Macdonald polynomial
also known as non-normalized Rogers polynomials. This is what we expected to obtain since by construction $P^{(-)}$ are eigenfunctions of 
Macdonald operator which can be directly obtained after taking $Q\to 0$
limit of \eqref{u2:rs:instanton}.

\subsection{$A_2$ RS from instantons.}
\label{sec:A2:insatnons}

All the arguments above are easily generalizable to the RS models defined on any $A_N$ root system. Since in our index 
computations presented in the Section \eqref{section:index} we have considered only $A_1$ and $A_2$ cases we will not go as far and 
will just summarize some details of $A_2$ model here.

In order to construct eigenfunctions of the $A_2$ RS model \eqref{A2:RS:action}  we start with the ramified instanton partition function 
$Z_{[1^3]}$ of $\SU(3)$ $\NN = 1^*$ $5d$ theory in the presence of the monodromy defect of type $[1,1,1] \equiv [1^3]$.
Now there are three 
instanton numbers $q_i$ corresponding to three sectors. In case of 
$\SU(2)$ theory we had two types of ramified instantons labeled by 
$\pm$ according to the parametrization of the two 
instanton numbers in \eqref{a1:instanton:parameters}.
Here in turn there are in total 3! choices according to 
embeddings of $\U(1)^3$ into $\U(3)$ labelled by permutations $\sigma$. However the results of different embeddings are the same up to 
permutations of the Coulomb branch parameters. Hence 
without loss of generality 
we further consider only one choice corresponding to the 
trivial permutation $\sigma =1$. From the point of view of the $3d$ $\NN = 4$ theory of a defect this choice corresponds to a specific choice of 
supersymmetric vacua. In this case instanton parameters can be chosen as 
follows: 
\be 
q_1 = z_1\,,\qquad q_2 = z_2\,,\qquad q_3 = Q z_3\,,
\label{A2:instanton:parameters}
\ee
where $z_1z_2z_3 = 1$ and $Q$ is the  usual $5d$ instanton parameter.

 As in $A_1$ case we are interested in the NS limit 
$\epsilon_2 \to 0$, which is a divergent limit. Once again to get rid of this divergence we normalize ramified instanton 
partition function as follows: 
\be
D_{[1^3]} = \lim_{\epsilon_2\to 0} \frac{Z_{[1^3]}}{Z}\,,
\ee
where $Z$ is the usual Nekrasov partition function without any defect. 
In this case partition function appears to be very complicated so we do not provide particular expression here. As was conjectured 
\cite{Nekrasov:2009rc,Bullimore:2014awa} these normalized ramified instanton partition function satisfy following equations:
\be
\left[\mu_1 \frac{\thfQ{\eta^2\frac{\tau_2}{\tau_1}}\thfQ{\eta^2\frac{\tau_3}{\tau_1}}}{\thfQ{\frac{\tau_2}{\tau_1}}\thfQ{\frac{\tau_3}{\tau_1}}}
p_{1} + \mu_2\eta^2 \frac{\thfQ{\eta^2\frac{\tau_1}{\tau_2}}\thfQ{\eta^2\frac{\tau_3}{\tau_2}}}{\thfQ{\frac{\tau_1}{\tau_2}}\thfQ{\frac{\tau_3}{\tau_2}}}
p_{2} +
\right. \nn\\\left.
\mu_3\eta^4 \frac{\thfQ{\eta^2\frac{\tau_1}{\tau_3}}\thfQ{\eta^2\frac{\tau_2}{\tau_3}}}{\thfQ{\frac{\tau_1}{\tau_3}}\thfQ{\frac{\tau_2}{\tau_3}}}
p_{3}\right]D_{[1^3]} = E_{(1)}D_{[1^3]}
\nn\\
\left[\mu_2\mu_3\eta^4 \frac{\thfQ{\eta^2\frac{\tau_1}{\tau_2}}\thfQ{\eta^2\frac{\tau_1}{\tau_3}}}{\thfQ{\frac{\tau_1}{\tau_2}}\thfQ{\frac{\tau_1}{\tau_3}}}
p_{1}^{-1} + \mu_3\mu_1\eta^2 \frac{\thfQ{\eta^2\frac{\tau_2}{\tau_3}}\thfQ{\eta^2\frac{\tau_2}{\tau_1}}}{\thfQ{\frac{\tau_2}{\tau_3}}\thfQ{\frac{\tau_2}{\tau_1}}}
p_{2}^{-1} +
\right. \nn\\\left.
\mu_1\mu_2 \frac{\thfQ{\eta^2\frac{\tau_3}{\tau_1}}\thfQ{\eta^2\frac{\tau_3}{\tau_2}}}{\thfQ{\frac{\tau_3}{\tau_1}}\thfQ{\frac{\tau_3}{\tau_2}}}
p_{3}^{-1}\right]D_{[1^3]} = E_{(1,1)}D_{[1^3]}
\label{A2:instanton:eqs}
\ee
Here $\mu_i$ are Coulomb branch parameters satisfying $\prod_{i=1}^3 \mu_i=1$ and $p_i$ are shift operators. 
If we considered ramified instanton partition functions in $\U(3)$ $\NN=1^*$ gauge theory this shift operators would be just shifting 
$\tau_i$ parameters, $p_i\tau_j = q^{\delta_{ij}}\tau_j$. However in the case of $\SU(3)$ gauge group $\tau_i$ parameters are not independent 
and shift operators $p_i$ act as follows:
\be
p_1:\quad \tau_1\to q^{2/3}\tau_1\,,\quad \tau_2\to q^{-1/3}\tau_2\,,\quad \tau_3\to q^{-1/3}\tau_3\,,
\nn\\
p_2:\quad \tau_1\to q^{-1/3}\tau_1\,,\quad \tau_2\to q^{2/3}\tau_2\,,\quad \tau_3\to q^{-1/3}\tau_3\,,
\nn\\
p_3:\quad \tau_1\to q^{-1/3}\tau_1\,,\quad \tau_2\to q^{-1/3}\tau_2\,,\quad \tau_3\to q^{2/3}\tau_3\,,
\label{su3:shift:operators}
\ee
Variables  $\tau_i$ constitute  more convenient parametrization 
of the instanton counting parameters $q_i$ introduced in \eqref{A2:instanton:parameters}. The relation to $z_i$ parametrization 
is as follows: 
\be
z_1 = \frac{\tau_2}{\tau_1}\,,\qquad z_2 = \frac{\tau_3}{\tau_2}\,,\qquad z_3 = \frac{\tau_1}{\tau_3}\,.
\ee
Eigenvalues $E_{(1)}$ and $E_{(1,1)}$ similarly to $A_1$ case have simple physical interpretations  in terms of defects 
in the gauge theory we consider. In particular $E_{(1)}$ and $E_{(1,1)}$ are expectation values of the Wilson loops in
fundamental and rank-two antisymmetric representations of the $\SU(3)$ gauge group  correspondingly.

Once again \eqref{A2:instanton:eqs} are not exactly $A_2$ RS Hamiltonians as given in \eqref{A2:RS:action}.
In order to obtain correct expressions one needs to introduce extra factor in front of the ramified instanton 
partition function. In this case the factor takes the following form:
\be
D(\tau_i) = \tau_1^{\log_q\left[\mu_1\eta^{-2}\right]}\tau_2^{\log_q\mu_2}\tau_3^{\log_q\left[\mu_3\eta^{2}\right]}D_{[1,1]}\,.
\label{A2:ramified:prefactor}
\ee
Then the function $D(\tau_i)$ defined in this way satisfies the following pair of equations: 
\be
\left[ \frac{\thfQ{\eta^2\frac{\tau_2}{\tau_1}}\thfQ{\eta^2\frac{\tau_3}{\tau_1}}}{\thfQ{\frac{\tau_2}{\tau_1}}\thfQ{\frac{\tau_3}{\tau_1}}}
p_{1} + \frac{\thfQ{\eta^2\frac{\tau_1}{\tau_2}}\thfQ{\eta^2\frac{\tau_3}{\tau_2}}}{\thfQ{\frac{\tau_1}{\tau_2}}\thfQ{\frac{\tau_3}{\tau_2}}}
p_{2} +
\right. \nn\\\left.
 \frac{\thfQ{\eta^2\frac{\tau_1}{\tau_3}}\thfQ{\eta^2\frac{\tau_2}{\tau_3}}}{\thfQ{\frac{\tau_1}{\tau_3}}\thfQ{\frac{\tau_2}{\tau_3}}}
p_{3}\right]D\left( \tau_i \right) = E_{(1)}D\left(\tau_i\right)
\nn\\
\left[\frac{\thfQ{\eta^2\frac{\tau_1}{\tau_2}}\thfQ{\eta^2\frac{\tau_1}{\tau_3}}}{\thfQ{\frac{\tau_1}{\tau_2}}\thfQ{\frac{\tau_1}{\tau_3}}}
p_{1}^{-1} +  \frac{\thfQ{\eta^2\frac{\tau_2}{\tau_3}}\thfQ{\eta^2\frac{\tau_2}{\tau_1}}}{\thfQ{\frac{\tau_2}{\tau_3}}\thfQ{\frac{\tau_2}{\tau_1}}}
p_{2}^{-1} +
\right. \nn\\\left.
 \frac{\thfQ{\eta^2\frac{\tau_3}{\tau_1}}\thfQ{\eta^2\frac{\tau_3}{\tau_2}}}{\thfQ{\frac{\tau_3}{\tau_1}}\thfQ{\frac{\tau_3}{\tau_2}}}
p_{3}^{-1}\right]D\left(\tau_i\right) = E_{(1,1)}D(\tau_i)
\label{A2:RS:action:instantons}
\ee
These equations in turn exactly describe the action of $A_2$ RS Hamiltonians. 	

Just as in the case of $A_1$ RS model ramified instanton partition functions are only formal eigenfunctions of 
the Hamiltonians specified in \eqref{A2:instanton:eqs} that depend on the set of Coulomb branch parameters $\mu_i$. 
In order to obtain the true spectrum of $A_2$ RS these Coulomb branch parameters should be quantized accordingly. 
In this case the proper $B$-model quantization condition should take the following form:
\be
\frac{\mu_{i+1}}{\mu_i} = \eta^2 q^{n_i}\,, \qquad  n_i \in \mathbb{Z}_{\geq 0}\,,\quad i=1,2\,,
\label{A2:B-model:quantization}
\ee
which is similar to the quantization in $A_1$ case with the choice of $+$ sign in  \eqref{B-model:quantization}.

\section{Matching ramified instantons and index calculations.}
\label{section:matching}

In the previous two sections we have summarized two ways of deriving 
the spectrum of RS integrable system. One approach presented in 
Section \ref{section:index} was proposed in \cite{Nazzal:2023wtw} and 
relies on the fact that superconformal indices of $4d$ $\NN=2$ class $\mathcal{S}$ theories 
are kernel functions of the RS Hamiltonians. Second approach uses ramified instanton 
partition functions of $5d$ $\NN=1^*$ SYM which are themselves eigenfunctions of RS Hamiltonians.
The latter approach was proposed in \cite{Nekrasov:2009rc} and directly checked in \cite{Bullimore:2014awa}
at the level of formal eigenfunctions. However the spectrum itself was first derived in \cite{Hatsuda:2018lnv}
where authors checked obtained eigenvalues against results of the numerical solution for the spectrum. 

Now when we have an alternative derivation of the spectrum from superconformal indices we can 
compare the two approaches directly for both eigenvalues and eigenfunctions of RS Hamiltonians which we 
do in the present section. 

In order to make a comparison of the two approaches first of all we need to establish the dictionary between 
index fugacities and parameters of the instanton counting. The easiest way to do it is to compare 
RS Hamiltonians in two cases. $A_1$ Hamiltonians \eqref{RS:hamiltonian:2} and \eqref{u2:rs:instanton}
is already  enough to find the right map which we summarize in the table below:

\begin{table}
\centering
\begin{tabular}{|c|c|}
\hline
Superconformal Indices & Ramified Instantons   \\ \hline
 $x_{i}$ & instanton parameters $\tau_i$  \\ \hline
 $t$    & adjoint mass $\eta^{2}$  \\ \hline
 $p$   & instanton parameter $Q$  \\ \hline
 $q$   & equivariant parameter $q=e^{\epsilon_1}$   \\\hline
\end{tabular}
\caption{Map between fugacities of superconformal indices and parameters of $5d$ $\NN=1^*$ 
gauge theory used in expression for the ramified partition functions.}
\label{dictionary}
\end{table}

One can check that this map also works for $A_2$ Hamiltonians as well. 
Using this dictionary we can directly compare eigenvalues and eigenfunctions of $RS$ models derived from superconformal 
index and ramified instanton calculations. 

\textbf{$A_1$ RS model.} In $A_1$ case index calculation provides us both eigenvalue  and eigenfunction for the ground state given in \eqref{A1RS:ground:state:energy} and 
\eqref{A1RS:eigfunct:result} correspondingly. Additionally from the superconformal index we extracted the wave function of the first excited state as given in 
\eqref{A1RS:eigfunct:result}. We can compare all these results with what one gets from the instanton calculus. 

First and easiest thing to do is to compare the ground state energy. In case of superconformal index it is given in \eqref{A1RS:ground:state:energy}. In case 
of the ramified instanton partition function it can be read from the eigenfunction equation \eqref{u2:rs:instanton} 
together with $\U(2)$ and $\U(1)$ Wilson loop expectation values given in \eqref{u2:wilson:loop} and \eqref{u1:wilson:loop}.
These expressions depend on the Coulomb branch parameters $\mu_{1,2}$. Using B-model quantization  
\eqref{B-model:quantization}  and applying parameters map specified in Table \ref{dictionary} 
we can obtain that the resulting ground state eigenvalues of $A_1$ RS model coincide in the two calculations. 

We should mention that eigenvalue obtained from ramified instanton counting with the B-model quantization \eqref{B-model:quantization} 
for the Coulomb branch parameters has already been compared against the numerical calculation of the corresponding eigenvalue of RS model 
in \cite{Hatsuda:2018lnv}. 
Here however we have even a more powerful check since we can compare eigenfunctions themselves with full $x$-dependencies. For 
superconformal index we can read corresponding expression directly from \eqref{A1RS:eigfunct:result}. In order to compare 
with results of instanton counting we should use parameters map from the Table \ref{dictionary} and apply Coulomb branch 
B-model quantization condition \eqref{B-model:quantization} with $n=0$ and $n=1$ for the ground and the first excited states correspondingly.
Notice that comparing the two approaches we should only take into account 
$x$-dependent parts since the two functions are expected to be the same up to normalization condition. In fact 
eigenfunctions obtained from superconformal index are normalized w.r.t. the integration measure \eqref{A1:index:measure} 
while the normalization of 5d partition functions on $\mathbb{R}^4$ depends on the boundary condition at infinity, which introduces an ambiguity that needs to be fixed. Because of this reason we can 
just check the ratio of the eigenfunction obtained in two approaches. Doing so we find for the ground state:
\be
\frac{\psi_0^{RS}(x)}{\left.D^{(-)}(x)\right|_{\mu_2=\mu_1^{-1}=t^{1/2} }} = 1+\left(t-\frac{2}{t}\right)\sqrt{pq} + \frac{1}{2}\sqrt{pq}(5p+3q)t - \frac{1}{2t}\sqrt{pq}(p+5q)
+ 
\nn\\
\frac{1}{2}pq\left(t^2 - 4 + \frac{5}{t^2} \right) + O\left( (pq)^{3/2} \right)\,.
\label{A1:ground:state:ratio}
\ee
For the first excited state we obtain: 
\be
\frac{\psi_1^{RS}(x)}{2\left.D^{(-)}(x)\right|_{\mu_2=\mu_1^{-1}=t^{1/2}q^{1/2}}} =
	1+(p+q) + \sqrt{pq} \left( 3t-\frac{1}{t}\right) +  \sqrt{pq}(p+q)\left( 6t-\frac{3}{t}\right)+
	\nn\\
	q\sqrt{pq}t + 3pq\left(2t^2 -1  \right) + 
	\left(p^2+q^2  \right) +  O\left( (pq)^{3/2} \right)\,.
	\label{A1:excited:ratio}
\ee
As we see in both cases the ratio does not depend on $x$ variables so they differ only by normalization. Here we present expansion up to an order 
$O\left(p^nq^m\right)$ with $n+m=2$. In our calculations we went up to an order with $n+m=3$ but we do not present corresponding results here 
since they are bulky and not very informative. Since $\psi_i^{RS}$ expansion is in $(pq)^{1/4}$ with $p/q$ fixed we
 effectively expand up to sixth order which is high enough 
to conclude that the ratios above are indeed independent of $x$ variables and both approaches give the same result for the RS spectrum. 

\

\noindent\textbf{$A_2$ RS model.} The same comparison can be performed for the  $A_2$ RS model. In this case however superconformal index calculation 
can only provide us the ground state wave function given in \eqref{A2:RS:ground:state}. Due to complexity of the calculation we are not capable 
of reproducing excited states and even a ground state eigenvalue since ground state wave function is not derived to a high enough order. 

We now can compare eigenfunctions \eqref{A2:RS:ground:state} with the ramified instanton partition functions of $\SU(3)$ $\NN=1^*$ gauge theory.
In particular we should consider function $D\left( \tau_i \right)$ as defined in \eqref{A2:ramified:prefactor}, impose B-model 
quantization \eqref{A2:B-model:quantization} with $n_1=n_2=0$ (ground state) and use parameters map in Table \ref{dictionary}.
Once again since in these two approaches results are expected to have different normalizations we should consider only $x$-dependent 
part. For this purpose we calculate the ratio of two functions and obtain: 
\be
\frac{\psi_{\emptyset}^{RS}(x_1,x_2,x_3)}{\left.D(x_1,x_2,x_3)\right|_{Eq.\eqref{A2:B-model:quantization}}} = 
	1 + 2\sqrt{pq}\left(t-\frac{2}{t}\right)+\frac{1}{2}\sqrt{pq}(7p+5q)t -
	\nn\\
	\frac{1}{2t}\sqrt{pq}(3p+7q) + 
	\frac{1}{2}pq\left(16+5t^2 + 13\frac{1}{t^2} \right) + O\left( (pq)^{3/2}  \right)
\label{A2:ground:state:ratio}
\ee
As can be seen the ratio we obtain is independent of $x$ variables RS Hamiltonian is acting on. Hence the ground state wave functions 
obtained in two approaches differ only by an overall normalization. Unlike in $A_1$ case here we were able to expand only up an 
order $O\left( p^nq^m \right)$ with $n+m = 2$ which we present here. However already at this order expansion strongly suggests that 
the eigenfunctions obtained in two approaches are the same.




\section{Instantons on $S_b^3\times\Sigma$ and relation to indices.}
\label{section:gluing}

In this section, we explore in more details the relationship between A-model quantization and B-model quantization, which are used in the main context, from the perspective of the 5-dimensional partition functions of the $\mathcal{N}=2$ SYM theories on $S_b^3\times\Sigma$. We start by reviewing the localization computation of these 5d partition functions with a topological A-twist investigated in \cite{Crichigno:2018adf}. We will then speculate on how to connect these localization results with the 4d index results derived from the B-model quantization, which is based on the modularity of the 4d superconformal index.

Let us consider a 5d $\mathcal{N}=2$ super Yang-Mills theory with an $SU(N)$ gauge group placed on a rigid background $S^3\times S^2$. On this background, we can turn on squashing parameters for both $S^3$ and $S^2$, as well as mass parameters for the flavor symmetries, while preserving a supercharge. This setup allows us to compute the partition function using the standard equivariant localization technique. 

The resulting partition function is given by a product of contributions from four fixed points, which are invariant points under the $U(1)\times U(1)$ isometry of $S_{\epsilon_1}^2\times S_{\epsilon_2}^2 \subset S_b^3 \times S_{\epsilon_2}^2$ background where the $S_b^3$ is an $S^1$ fibration over the first $S_{\epsilon_1}^2$. Here $b$ is the squashing parameter of $S^3$ while $\epsilon_{1,2}$ are equivariant parameters on 
two $S^2$.
Moreover, the local geometry near each fixed point can be approximated as $S^1\times \mathbb{R}^4$. Therefore, the partition function is expressed as the product of four copies of the 5d Nekrasov partition functions on an $\Omega$-deformed $S^1\times \mathbb{R}^4$ background and can be written as \cite{Crichigno:2018adf}
\begin{align}\label{eq:full-function}
	Z_{S^3 \times S^2}(m,\tau,\epsilon_{1,2})_{\mathfrak n} = \sum_{\mathfrak m} \oint da \prod_{\ell} Z_{\mathbb{R}^2_{\epsilon^{(\ell)}_1}\times
 \mathbb{R}^2_{\epsilon^{(\ell)}_1}}(a^{(\ell)},m^{(\ell)},\tau^{(\ell)};\epsilon_1^{(\ell)},\epsilon_2^{(\ell)}) \ ,
\end{align}
where $Z_{\mathbb{R}^4}$ is the 5d Nekrasov partition function on $S^1\times \mathbb{R}^4$ which involves all classical, perturbative, and instanton contributions. The variables $a,m,\tau$ correspond to the Coulomb branch parameters, the adjoint mass parameter, and the gauge coupling (or the instanton counting parameter $Q=e^{2\pi i\tau}$), respectively, and ${\mathfrak m}, {\mathfrak n}$ are the magnetic fluxes on (second) $S^2$ associated with the $SU(N)$ gauge symmetry and the $SU(2)_m$ flavor symmetry, respectively. The index $\ell$, which takes the values $\{nn,ns,sn,ss\}$ denotes one of four fixed points located at the north ($n$) and south ($s$) poles of the two spheres in this background. Parameters 
$\epsilon_{1,2}^{(\ell)}$ of the $\Omega$-deformation at each of  the fixed points are given by \cite{Crichigno:2018adf}:
\begin{align}
	&\epsilon^{(\ell)}_1 = \left\{\begin{array}{cc}b^2\,, & \!\! \ell = nn\ {\rm or} \ ns \\  b^{-2}, & \!\!\ell = sn\ {\rm or} \ ss\end{array}\right., \quad \epsilon^{(\ell)}_2 = \left\{\begin{array}{cc}\epsilon_2\,, & \!\!\ell = nn\ {\rm or} \ sn \\ -\epsilon_2\,, &\!\! \ell = ns\ {\rm or} \ ss\end{array} \right., \quad \tau^{(\ell)}= \left\{\begin{array}{cc}b^{-1}\gamma^{-1}\,, & \!\!\ell = nn\ {\rm or} \ sn \\ b\gamma^{-1}\,, & \!\!\ell = ns\ {\rm or} \ ss\end{array} \right., \nonumber \\
    &a^{(\ell)} = \left\{\begin{array}{cc}b\gamma a + \tfrac{{\mathfrak m}}{2}\epsilon_2^{(\ell)}\,, & \ell = nn\ {\rm or} \ ns \\  b^{-1}\gamma  a + \tfrac{{\mathfrak m}}{2}\epsilon_2^{(\ell)}, & \ell = sn\ {\rm or} \ ss \end{array}\right. , \quad m^{(\ell)} = \left\{\begin{array}{cc}b\gamma m +\tfrac{1}{2}+ \tfrac{{\mathfrak m}}{2}\epsilon_2^{(\ell)}\,, & \ell = nn\ {\rm or} \ ns \\  b^{-1}\gamma m  +\tfrac{1}{2}+ \tfrac{{\mathfrak m}}{2}\epsilon_2^{(\ell)}, & \ell = sn\ {\rm or} \ ss \end{array}\right.\,,
\end{align}
Now, we perform a topological A-twist which can be achieved by taking the limit $\epsilon_2\rightarrow0$ in the partition function. Then the local partition function at each fixed point reduces to the Nekrasov-Shatashvili (NS) limit of the 5d partition function which can be written as
\begin{align}
	Z_{\mathbb{R}^2_{\epsilon_1}\times \mathbb{R}^2_{\epsilon_2}}(a,m,\tau) \ \ \stackrel{\epsilon_2\rightarrow0}{\longrightarrow} \ \ {\rm exp}\left[2\pi i \left(\frac{1}{\epsilon_2}\mathcal{W}_{NS}(a,m,\tau,\epsilon_1) - \frac{1}{2}\Omega_{NS}(a,m,\tau,\epsilon_1)\right) + \mathcal{O}(\epsilon_2)\right] \ ,
\end{align}
where $\mathcal{W}_{NS}$ and $\Omega_{NS}$ are called the {\it effective twisted superpotential} and the {\it effective dilaton}, respectively, characterizing the  effective 2d theory on the $\mathbb{R}^2_{\epsilon_2}$ plane at low energy. We can collect the partition functions in the NS-limit at four fixed points and rewrite them in the following manner 
\begin{align}
	Z_{S^3_b\times S^2}(m,\tau)_{\mathfrak n} = \frac{1}{N!}\sum_{\mathfrak m}\oint da\, \Pi_i(a,m,\tau)^{\mathfrak m} \Pi_m(a,m,\tau)^{\mathfrak n} e^{-2\pi i\Omega_{S_b^3\times\mathbb{R}^2}(a,m,\tau)} \ ,
 \label{S3xS2:product:decompose}
\end{align}
where
\begin{align}
	\Pi_i(a,m,\tau) = e^{2\pi i\partial_{a_i}\mathcal{W}_{S_b^3\times\mathbb{R}^2}(a,m,\tau)} \ , \quad \Pi_m(a,m,\tau) = e^{2\pi i\partial_{m}\mathcal{W}_{S_b^3\times\mathbb{R}^2}(a,m,\tau)} \ .
\end{align}
Here, $\mathcal{W}_{S_b^3\times \mathbb{R}^2}$ and $\Omega_{S_b^3\times \mathbb{R}^2}$ represent the sum of the contributions to the effective twisted superpotential and the effective dilaton, respectively, from the north and south poles of $S_{\epsilon_1}^2\subset S_b^3$.
The detailed expressions for these functions can be found in \cite{Crichigno:2018adf}.

The integration should be performed carefully using appropriate contours. The correct contour integral prescription has been studied in \cite{Crichigno:2018adf}, which extends the prescription described in \cite{Closset:2017zgf} for 3d partition functions with a topological twist. By employing this contour prescription, the partition function reduces to the form of the Bethe-vacua summation formula:
\begin{align}
	Z_{S^3_b\times S^2}(m,\tau)_{\mathfrak n} = \sum_{\hat{a}\in \mathcal{S}_{BE}}\Pi_m(\hat{a},m,\tau)^{\mathfrak n}\,\mathcal{H}(\hat{a},m,\tau)^{-1} \ ,
\end{align}
where $\mathcal{S}_{BE}$ is the set of ``Bethe vacua'' defined as the solutions to the Bethe equations, which are given by
\begin{align}\label{eq:Bethe}
	\mathcal{S}_{BE} = \left\{\hat{a}\ \Big|\ {\rm exp}\left(2\pi i \frac{\partial\mathcal{W}_{S_b^3\times \mathbb{R}^2}(\hat{a},m,\tau)}{\partial a_i}\right)=1 ,\ i=1,\cdots,N-1\right\}/W_G \ ,
\end{align}
and
\begin{align}
	\mathcal{H}(\hat{a},m,\tau) = e^{2\pi i\Omega_{S_b^3\times\mathbb{R}^2}(a,m,\tau)}{\rm det}\left(\partial_{a_i}\partial_{a_j}\mathcal{W}_{S_b^3\times \mathbb{R}^2}(a,m,\tau)\right) \ .
\end{align}
The Bethe vacua for $\mathcal{S}_{BE}$ should be divided by the action $W_G$ of the $SU(N)$ Weyl group, and the solutions that are not acted freely by the Weyl group need to be excluded. We remark that this set $\mathcal{S}_{BE}$ is precisely the solutions to the A-model quantization condition. The connection between the A-model quantization condition and 5d partition functions in NS-limits on other compact manifolds has been discussed in \cite{Sciarappa:2017hds,Hatsuda:2018lnv}.

The partition function with a co-dimension two defect can be calculated similarly. Consider a co-dimension two defect inserted at the north pole on $S^2$ which we will twist later. This will affect two out of four fixed points $\ell=nn$ and $\ell = sn$ contributing to the 
partition function \eqref{S3xS2:product:decompose}. These two points correspond to the partition function on $S_b^3\times \mathbb{R}^2$ 
around the north pole on the twisted $S^2$. For these points we have to substitute usual Nekrasov partition functions 
in \eqref{S3xS2:product:decompose} with the ramified instanton partition functions discussed in Section \ref{sec:instantons}. 
 Then we take the NS-limit, i.e. $\epsilon_2\rightarrow 0$, to implement an A-twist on $S^2$.  In this limit, the defect's contribution to the partition function remains finite because it is localized at a point on $S^2$ and therefore does not involve the divergent factor $1/\epsilon_2$.  Moreover, this contribution does not depend on the magnetic fluxes ${\mathfrak m,\mathfrak n}$ on $S^2$ due to the localized nature of the defect, preventing it from capturing the overall flux profile. Therefore, the full partition function with the defect can be written as
\begin{align}\label{eq:S3S2-defect}
	Z_{S^3_b\times S^2}^{\rm def}(m,\tau)_{\mathfrak n} = \frac{1}{N!}\sum_{\mathfrak m}\oint da\, \Pi_i(a,m,\tau)^{\mathfrak m} \Pi_m(a,m,\tau)^{\mathfrak n} e^{-2\pi i\Omega_{S_b^3\times\mathbb{R}^2}(a,m,\tau)}\, \psi(x,a,m,\tau) \ ,
\end{align}
where
\begin{align}
	\psi(x,a,m,\tau) = \lim_{\epsilon_2\rightarrow 0} \prod_{\ell=nn,sn}\frac{Z^{\rm def}_{\mathbb{R}^4}(x,a^{(\ell)},m^{(\ell)},\tau^{(\ell)},\epsilon_1^{(\ell)},\epsilon_2^{(\ell)})}{Z_{\mathbb{R}^4}(a^{(\ell)},m^{(\ell)},\tau^{(\ell)},\epsilon_1^{(\ell)},\epsilon_2^{(\ell)})} \ ,
\end{align}
where $Z^{\rm def}_{\mathbb{R}^4}$ is the ramified instanton partition function on $\mathbb{R}^4$ with defect parameter $x$.

The contribution from the defect in the integrand is subleading in the $\epsilon_2$ expansion and thus it does not alter the pole structure in the integration.  Therefore, the contour integration proceeds as previously, and the resulting partition function can again be expressed as a sum over the same Bethe vacua as before:
\begin{align}\label{eq:Bethe-sum-defect}
	Z_{S^3_b\times S^2}^{\rm def}(x,m,\tau)_{\mathfrak n} = \sum_{\hat{a}\in \mathcal{S}_{BE}}\Pi_m(\hat{a},m,\tau)^{\mathfrak n}\,\mathcal{H}(\hat{a},m,\tau)^{-1}\, \psi(x,\hat{a},m,\tau) \ .
\end{align}

From the perspective of the 4d index, this computation yields the superconformal index of the 4d class $\mathcal{S}$ theory arising from 6d $A_{N-1}$ (2,0) SCFT compactified on a sphere with a (maximal) puncture. Here, the puncture comes from the co-dimension two defect, and its effect is encoded in $\psi(x,\hat{a},m,\tau)$ in the localization computation. Thus, this approach offers a method to calculate the superconformal index contribution from the punctures in 4d class $\mathcal{S}$ theories using the 5d maximal SYM theories.

However, this computation, which we call the A-model computation, is carried out under a different parameter regime from that typically used in the 4d superconformal index.  In fact, the above partition function is given by an instanton expansion, which holds when the 5d gauge coupling $\tau$ is small. In contrast, the 4d index is typically expanded using fugacities when $\tau$ is large. Therefore, the results from these two different parameter regime cannot be compared immediately. It turns out that in order to accurately translate the defect contributions from the 5d context into the 4d index, we need an S-duality transformation, which converts $\tau\rightarrow -\frac{1}{\tau}$. 

The modular properties of supersymmetric partition functions of 4d $\mathcal{N}=1$ superconformal field theories have been extensively studied in \cite{Gadde:2020bov}, which generalizes the modular properties of elliptic genera of two-dimensional theories. Note that the 4d superconformal index can be obtained from a supersymmetric partition function on an $S^1\times S^3$ background and this background can be formed by gluing two solid three-tori $\mathbb{T}^2\times D^2$, where an action of the S generator in $SL(2,\mathbb{Z})\in SL(3,\mathbb{Z})$ is applied to a boundary $\mathbb{T}^3$ in the gluing process. In this setup, the large diffeomorphism group $SL(3,\mathbb{Z})$ acts projectively on the complex structures, represented by the column vector $\vec{\tau}\equiv (1,\sigma,\tau)^T$, of the boundary $\mathbb{T}^3$. 

The investigation in \cite{Gadde:2020bov} suggests that the consistency of this gluing procedure under the choice of a large diffeomorphism of the boundary torus and the choice of a large gauge transformation for the background flavor symmetry, which takes an element in a semi-direct product group $SL(3,\mathbb{Z})\times (\mathbb{Z}^3)^r$ with rank $r$ flavor symmetry, leads to a non-trivial modular property of the superconformal index:
\begin{align}\label{eq:modular}
	Z^\alpha_{Y}(\vec\tau) Z^\alpha_{Y}(Y^{-1}\cdot \vec\tau) Z^\alpha_{Y}(Y^{-2}\cdot \vec\tau) = e^{-i\frac{\pi}{3}P(\vec\tau)} \ ,
\end{align}
where $Y\in SL(3,\mathbb{Z})$ cyclically permutes $(1,\sigma,\tau)$.  The phase factor $P(\vec\tau)$ is the 't-Hooft anomaly polynomial of the 4d theory. Here, the index $\alpha$ stands for a supersymmetric vacuum in the 1d quantum mechanics defined along the time circle $S^1$. In our notation for the superconformal index, the complex structures $\sigma,\tau$ are related to the fugacities as $q=e^{2\pi i \sigma}, p=e^{2\pi i\tau}$, and a supersymmetric vacuum labelled by $\alpha$ corresponds to a Bethe vacuum $\hat{a}\in \mathcal{S}_{BE}$.
The modular property of the 4d superconformal index can provide an S-duality transformation that we need for the relationship between the 5d partition function on $S_b^3\times S^2$ and the 4d index of class $\mathcal{S}$ theories. Let us illustrate this with a simple but concrete example. 

Consider the 5d $\mathcal{N}=2$ $U(1)$ gauge theory which arises from a circle compactification of the 6d $(2,0)$ theory on a single M5-brane. The 5d partition function of this theory on the $\Omega$-deformed background $S^1\times \mathbb{R}^4_{\epsilon_{1,2}}$ was computed in \cite{Nekrasov:2002qd,Nekrasov:2003rj,Iqbal:2008ra,Kim:2011mv}. The results is
\begin{align}
    Z_{\mathbb{R}^2_{\epsilon_1}\times \mathbb{R}^2_{\epsilon_2}}(a,m,\tau)&=Z_{\mathbb{R}^2_{\epsilon_1}\times \mathbb{R}^2_{\epsilon_2}}^{\rm pert}(a,m,\tau)\times Z_{\mathbb{R}^2_{\epsilon_1}\times \mathbb{R}^2_{\epsilon_2}}^{\rm inst}(a,m,\tau) \nonumber \\
    Z_{\mathbb{R}^2_{\epsilon_1}\times \mathbb{R}^2_{\epsilon_2}}^{\rm pert}(a,m,\tau) &= {\rm PE}\left[\frac{-e^{4\pi i \epsilon_+} + e^{2\pi i \epsilon_++2\pi i m}}{(1-e^{2\pi i\epsilon_1})(1-e^{2\pi i\epsilon_2})}\right] \ ,
\end{align}
where $Z_{\mathbb{R}^2_{\epsilon_1}\times \mathbb{R}^2_{\epsilon_2}}^{\rm pert}$ is the perturbative contribution and $Z_{\mathbb{R}^2_{\epsilon_1}\times \mathbb{R}^2_{\epsilon_2}}^{\rm inst}$ is the instanton contribution. Function $\mathrm{PE}[e^a]$ stands for the plethystic exponential of a letter index $e^a$ and is given by the following expression: 
\be
\mathrm{PE}\left[f(e^a)\right]\equiv\mathrm{exp}\left[\sum_{n=1}^\infty \frac{f(e^{na})}{n}\right]
\label{plethistic:exponent}
\ee
The instanton contribution for the $U(1)$ theory can be computed exactly using the topological vertex method, as described in \cite{Iqbal:2008ra,Kim:2011mv}, and the result reduces to the simple expression as
\begin{align}
    Z_{\mathbb{R}^2_{\epsilon_1}\times \mathbb{R}^2_{\epsilon_2}}^{\rm inst}(a,m,\tau) ={\rm PE}\left[\frac{e^{2\pi i \tau^{-1}}}{1-e^{2\pi i \tau^{-1}}}\frac{e^{2\pi i (\epsilon_+-m)}(1-e^{2\pi i(m+\epsilon_-) })(1-e^{2\pi i(m-\epsilon_-) })}{(1-e^{2\pi i\epsilon_1})(1-e^{2\pi i\epsilon_2})}\right] \ .
\end{align}
 The partition function on $S_b^3\times S^2$ then can be computed by plugging this result into the expression in \eqref{eq:full-function}, and taking $\epsilon_2\rightarrow0$ limit. The parameters in the usual instanton partition function are mapped to those in the localized partition functions on $S_b^3\times S^2$ as follows \cite{Crichigno:2018adf}:
\begin{align}
    (nn),\,(ns) \ &: \ \epsilon_1 \rightarrow b^2 \ , \quad m \rightarrow b\gamma m +\frac{1}{2} \ , \quad \tau \rightarrow \tau\equiv b^{-1}\gamma^{-1} \ ,\nonumber \\
    (sn),\,(ss) \ &: \ \epsilon_1 \rightarrow b^{-2} \ , \quad m \rightarrow b^{-1}\gamma m +\frac{1}{2} \ , \quad \tau \rightarrow \sigma \equiv b\gamma^{-1} \ .
\end{align}
Here, the shift of the flavor mass parameter by $1/2$ is due to the non-trivial $S^1$ fibration within $S_b^3$ and this shift is essential for achieving the correct result \footnote{A similar shift of the flavor mass parameter due to a non-trivial fibration structure was also observed in the localization computation for the $S^5$ partition function in \cite{Kim:2012qf}}. 
One can then compute
\begin{align}
	Z_{S^3_b\times S^2}^{U(1)}(m,\tau)_{\mathfrak n} &= {\rm PE}\left[\frac{1-e^{2\pi i(b^2+b\gamma)}}{(1\!-\!e^{2\pi i b^2})(1\!-\!e^{2\pi ib\gamma})}-1\right.\\
     &\left.+ \frac{{\mathfrak n}\,e^{\pi i b^2}\!\left(e^{\pi i (2b\gamma-2b\gamma m-\frac{1}{2})}\!-\!e^{\pi i (2b\gamma m+\frac{1}{2})}\right)}{(1-e^{2\pi i b^2})(1-e^{2\pi ib\gamma})} + (b\rightarrow b^{-1})\right] \nonumber \\
	&= \Gamma(0,\frac{\sigma}{\tau},\frac{1}{\tau})\Gamma(0,\frac{\tau}{\sigma},\frac{1}{\sigma})\times \left(\Gamma(\frac{\hat{m}}{\tau},\frac{\sigma}{\tau},\frac{1}{\tau})\Gamma(\frac{\hat{m}}{\sigma},\frac{\tau}{\sigma},\frac{1}{\sigma})\right)^{-\mathfrak n} \nonumber \\
	&= \Gamma(0,\sigma,\tau)^{-1}\times \Gamma(\hat{m},\sigma,\tau)^{\mathfrak n}\ , 
\end{align}
where $\hat{m} \equiv m + \frac{\sigma+\tau}{2}$, up to a constant phase factor. For the last equality, we used the modular property of the elliptic gamma function $\Gamma$,
\begin{align}
	& \Gamma(z,\sigma,\tau) = \prod_{j,k=0}^\infty\frac{(1-e^{2\pi i (-z+(j+1)\sigma+(k+1)\tau)})}{(1-e^{2\pi i (z+j\sigma+k\tau)})}\ ,\nonumber \\
	&\Gamma(z,\tau,\sigma)\Gamma(\frac{z}{\tau},\frac{\sigma}{\tau},\frac{1}{\tau})\Gamma(\frac{z}{\sigma},\frac{1}{\sigma},\frac{\tau}{\sigma}) = e^{-i\frac{\pi}{3}Q(z,\tau,\sigma)} \ .
\end{align}

The final partition function is now expressed as a 4d superconformal index. It consists of contributions from a 4d $\mathcal{N}=1$ vector multiplet and ${\mathfrak n}$ chiral multiplet. This result perfectly agrees with the expected result for the 4d theory coming from a 6d $(2,0)$ free tensor multiplet compactified on a sphere with an $SU(2)_m$ flux $\mathfrak{n}$.\footnote{As $S^2$ has an $SU(2)$ isometry the $4d$ theory has an additional $SU(2)$ global symmetry. In particular  the chiral superfields will form the ${\mathfrak n}$ dimensional representation and we can refine further the index turning on a fugacity for the $SU(2)$ symmetry. See for example \cite{Hosseini:2020vgl,Hwang:2021xyw}.} Crucially, this expression for the 4d index was obtained by utilizing the modular property of the elliptic gamma function. This provides a clear and simple demonstration of the modular properties of the superconformal indices for 4d class 
$\mathcal{S}$ theories, given in \eqref{eq:modular}, and their realizations in the 5d partition functions.

It is natural to interpret the modular property in \eqref{eq:modular} as an S-duality transformation that inverts $\tau, \sigma$, which are proportional to the 5d gauge coupling, into their inverses $1/\tau$ and $1/\sigma$. We expect that this is also the case for other interacting class $\mathcal{S}$ theories beyond the Abelian example. Then, as also suggested in \cite{Hatsuda:2018lnv}, the Bethe vacua equation in \eqref{eq:Bethe} under this S-duality can be reformulated as
\begin{align}\label{eq:Bethe-dual}
    e^{2\pi i \frac{\partial\mathcal{W}}{\partial a_i}} = 1 \quad \rightarrow \quad e^{2\pi i a^D_i} = 1 \ \ {\rm with} \ \ a_i^D = \frac{\partial\mathcal{W}}{\partial a_i} \ ,
\end{align}
using the Seiberg-Witten relation, which is generalized by the squashing parameter $b$ and flavor mass parameters $z$, between the Coulomb branch parameters $a_i$ and their S-dual parameters $a_i^D$. We interpret this S-dualized Bethe vacua equation as the B-model quantization condition. So, we propose that the A-model and the B-model quantization conditions are related through the modular property of the 4d superconformal index and the corresponding 5d (or 6d) partition functions.

A concrete example supporting this proposal is the computation of the eigenfunction $\psi_\lambda$ of the RS model using the partition function of the 5d $\mathcal{N}=2$ gauge theories in the presence of the defects, which was presented in the previous sections. According to \cite{Gaiotto:2009we,Gaiotto:2009hg}, the punctures on the Riemann surface $\Sigma$ for class $\mathcal{S}$ theories of algebra $G$ correspond to the Gukov-Witten type monodromy defects \cite{Gukov:2006jk}, which are labeled by homomorphisms $\rho\ : SU(2) \rightarrow G$. The defect discussed in the main context and also in \eqref{eq:S3S2-defect} is exactly this monodromy defect of maximal type $\rho=[1^N]$ for $G=SU(N)$. Therefore, the puncture contribution to the 4d index of the class $\mathcal{S}$ theory is captured by the function $\psi$ in \eqref{eq:S3S2-defect} and \eqref{eq:Bethe-sum-defect}. Moreover, the modular property given in \eqref{eq:modular} of the 4d index implies that the summand at each Bethe vacuum in \eqref{eq:Bethe-sum-defect} enjoys a modular transformation that permutes complex structures $(1,\sigma,\tau)$ in the superconformal index. This should hold both with and without the defect. 

We therefore suggest that the defect contribution $\psi(\hat{a})$ itself should exhibit the modular property of the same form given in \eqref{eq:modular}. The localization computation of the 5d partition function on $S_b^3\times S^2$ with the defect includes contributions from a co-dim 2 defect inserted at a point on $S^2$. Schematically, this contribution can be expressed as a product of two local contributions, $\psi^N$ and $\psi^S$, at the north and south poles of $S^2\subset S_b^3$. Under the modular S-duality transformation in \eqref{eq:modular}, these two contributions are converted into a single function $\psi^D$, now expressed using dual variables $a_i^D$ evaluated at each Bethe vacuum. This dual function $\psi^D(\hat{a}_i^D)$ with the solution $\hat{a}_i^D$ to the Bethe equation \eqref{eq:Bethe-dual} computes the puncture contribution to the 4d index at each Bethe vacuum. Our explicit checks confirm that they match precisely with the eigenfunctions of the RS model, when each Bethe vacuum (or $n$) is identified with the energy level $\lambda$ of the RS Hamiltonian. Therefore, this provides a strong evidence for the modular properties of the partition functions of class $\mathcal{S}$ theories and the S-dual relation between the A-model and B-model quantization conditions.

\section{Discussion and Outlook}
\label{section:discussion}

In our paper we have investigated two approaches to the description of the 
spectrum of $A_N$ RS model. Both approaches use emergence of 
RS model in two different contexts of supersymmetric gauge theories. First 
approach recently proposed in \cite{Nazzal:2023wtw} relies on the 
computations of the suprconformal indices of $4d$ class $\mathcal{S}$ 
theories in the presence of surface defects.
In the second approach we calculate ramified instanton 
partition functions and Wilson loop expectation values
in $5d$ $\NN = 1^*$ $\SU(N)$ theories 
which constitute formal eigenfunctions and eigenvalues of 
$A_N$ RS model correspondingly \cite{Bullimore:2014awa}.

An important issue arising in ramified instanton calculations is the quantization of the Coulomb branch parameters, which turns formal eigenfunctions into the actual spectrum of the model. Proposals for the exact quantization conditions, referred to as A-model and B-model quantization conditions, are discussed in \cite{Hatsuda:2018lnv} and were validated against the numerical eigenvalues of the RS model. The A-model quantization can be understood through the saddle point equations of the partition functions on a compact manifold, as discussed in various studies, for example, in \cite{Sciarappa:2017hds,Crichigno:2018adf}. For the B-model quantization, we relate it to the A-model quantization via an S-duality transformation, which naturally acts on the 4d superconformal index \cite{Gadde:2020bov}. We demonstrated a concrete application of this for the 5d free $U(1)$ gauge theory, but detailed derivations of this S-duality transformation and the B-model quantization would be invaluable for extending this approach to other theories.

Finally we  compared results of the two approaches (index and instanton calculations) and matched corresponding ground states 
in cases of $A_1$ and $A_2$ RS models. For this we used B-model quantization condition proposed in \cite{Hatsuda:2018lnv}. 
We also discussed relation to another, A-model, quantization condition and its interpretation 
from the point of view of index calculations. 

There are still many questions left to be answered and there are many directions in which our research can be expanded. First of all in the 
present paper we have only analyzed spectrum of RS model. However methods we used in our paper can be easily extended to 
other integrable elliptic models. 

The first natural candidate for this investigation is van Diejen model \cite{vanDiejen, Noumi} 
which is $BC_1$ deformation of the RS model. Spectrum of van Diejen model is even less studied than the one of RS since not much 
is known even about its limits. From the point of view of indices it is related to the compactifications of $6d$ E-string theory 
on Riemann surfaces with punctures \cite{Nazzal:2018brc, Nazzal:2021tiu}. In our previous paper \cite{Nazzal:2023wtw} we have already made first steps towards 
describing ground state of  this model in certain regime of parameters. Some preliminary results have also been obtained in the context 
of instanton calculations \cite{Chen:2021ivd,Chen:2021rek}. It would be interesting to perform more in-depth detailed analysis of 
ground state and possibly higher states of van Diejen Hamiltonian using both of the approaches discussed in out paper. 

Also there exist many less studied elliptic integrable systems for which we can use both approaches discussed in the present 
paper. In fact already in our previous paper \cite{Nazzal:2023wtw} we have already derived ground state wavefunction and 
energy for the novel models defined on $A_2$ and $A_3$ root systems that were introduced in \cite{Razamat:2018zel, Ruijsenaars:2020shk}. These integrable 
systems are associated with $6d$ $\SU(3)$ minimal SCFT \cite{Seiberg:1996qx,Bershadsky:1997sb} and $\SO(8)$ minimal conformal matter theories correspondingly. It would be interesting to perform ramified instanton calculations and extend our analysis to these theories as well. 

Finally there are many more elliptic integrable models associated to $4d$ compactifications of various $6d$ SCFTs. Several novel systems 
corresponding to generalizations of van Diejen model on $A_N$ and $C_N$ root lattices were derived very recently
\cite{Nazzal:2021tiu,Nazzal:2023bzu,Nedelin:2023syx}. Many other examples of known and previously unknown integrable 
finite difference operators can be found in the literature \cite{Gaiotto:2015usa,Razamat:2013jxa,Lemos:2012ph}. It would be interesting to 
study spectra of these elliptic integrable systems using both superconformal index and ramified instanton approaches. This will 
help us to classify such systems and to better understand their general features. 

Another interesting direction for the future research is application of the obtained spectra to calculations of superconformal indices. 
Of course indices of the Lagrangian theories can be relatively easily calculated with conventional methods. 
However many $4d$ theories with $6d$ origin lack Lagrangian description which makes usual computational methods 
useless. In this case the spectrum of the corresponding integrable system associated to a given $4d$ theory 
can be used to effectively evaluate its superconformal index.
Such approach proved to be useful for class $\mathcal{S}$ theories \cite{Gadde:2009kb, Gadde:2011ik,Gadde:2011uv}
in certain limits (Macdonald, Schur etc.) where the spectrum of RS is known precisely and eigenfunctions take simple 
polynomial form. Now when we know the way for the derivation of RS spectrum in full without taking any limits we can extend this 
analysis of class $\mathcal{S}$ theories and also perform similar analysis for other theories discussed above.

\vspace{1cm}
\noindent{\bf Acknowledgments}:~
We are grateful to Alba Grassi for useful discussions. 
HK is supported by Samsung Science and Technology Foundation under Project Number SSTF-BA2002-05 and by the National Research Foundation of Korea (NRF) grant funded by the Korea government (MSIT) (2023R1A2C1006542). The work of HK at Harvard University is supported in part by the Bershadsky Distinguished Visiting Fellowship.
The research of SSR is supported in part by Israel Science Foundation under grant no. 2159/22, and by the Israeli Planning and Budgeting Committee. The research of AN is supported by STFC Grant 
No. ST/X000753/1 and at the initial stage of the project by the Swiss National Science Foundation, Grant No. 185723.

\appendix

\section{$A_1$ Macdonald polynomials}
\label{app:A1:Macdonald}

In this section we briefly review $A_1$ Macdonald polynomials. 
These polynomials are eigenfunctions of the following finite difference operator:
\be
M^{(q,t)}(x) = \frac{1-tx^2}{1-x^2}\Delta_{x\to q^{1/2}x}+\frac{1-tx^{-1}}{1-x^{-1}}\Delta_{x\to q^{-1/2} x} 
\label{A1:Macdonald:operator}
\ee
It can be easily seen that the operator on the l.h.s. of this equation is the Macdonald limit of the $A_1$  RS operator 
\eqref{RS:hamiltonian:2}. This limit is usually defined as taking $p\to 0$. However if we use notations of 
\eqref{RS:hamiltonian:2} we should first shift $t$ parameter as $t\to t/\sqrt{pq}$ and only then take $p\to 0$ limit. 
Performing this operation we will directly arrive to the Macdonald operator in \eqref{A1:Macdonald:operator}. 

Eigenfunctions of the $A_1$ Macdonald operator \eqref{A1:Macdonald:operator} are known as
\textit{$q$-ultraspherical polynomials} or \textit{Rogers polynomials} and are given by the following explicit expression:
\be
R_n\left(x;t,q \right)=\sum\limits_{k=0}^n\frac{(t;q)_k(t;q)_{n-k}}{(q;q)_k(q;q)_{n-k}}x^{2k-n}\,.
\label{rogers:polynom}
\ee
As eigenfunctions they  satisfy the following simple equations: 
\be
M^{(q,t)}(x) R_n\left(x;t,q \right)=
\left(q^{-\frac{n}{2}}+tq^{\frac{n}{2}}\right)R_n\left(x;t,q \right)
\label{A1:Macdonald:spectrum}
\ee
As defined here the polynomials are not normalized in any way. Natural normalization should be performed with
respect to the following integration measure:
\be
\Delta_{M}(x)=\frac{1}{2}\frac{\qPoc{x^{\pm 2}}{q}}{\qPoc{tx^{\pm 2} }{q}}\,,
\label{A1:macdonald:measure}
\ee
Then normalized properly eigenfunctions take the following form: 
\be
P_n(x;t,q)=\left[\frac{1-tq^n}{1-t}\frac{(q;q)_n\qPoc{q}{q}\qPoc{t^2}{q} }{(t^2;q)_n \qPoc{tq}{q}\qPoc{t}{q} }  \right]^{1/2}R_n(x;t,q)
\label{rogers:polynom:norm}
\ee
with the orthonormality condition 
\be
\oint\frac{dx}{4\pi i x}\Delta_M(x)P_n(x;t,q)P_m(x;t,q) = \delta_{nm}\,,
\label{macdonald:orthonormality}
\ee
which can be checked directly using \eqref{rogers:polynom:norm} and \eqref{A1:macdonald:measure}. 

Since in our computations we mainly consider ground state and first excited level it would be helpful to write down explicit expressions 
for these two eigenfunctions. In particular for the normalized Rogers polynomials \eqref{rogers:polynom:norm} for these two levels are given by 
\be
P_0(x;t,q)&=&\frac{\sqrt{(1-t)\qPoc{q}{q}\qPoc{t^2}{q}}}{\qPoc{t}{q}}\,,\nn\\
P_1(x;t,q)&=&\sqrt{\frac{(1-qt)(1-t)}{(1-q)(1+t)}\qPoc{q}{q}\qPoc{t^2}{q}}\frac{1}{\qPoc{t}{q}}\left(x+\frac{1}{x}\right)\,.
\label{A1:macdonald:levels12}
\ee
And the corresponding eigenvalues are given according to \eqref{A1:Macdonald:spectrum} by 
\be
E_0=1+t\,.\qquad E_1 = q^{-1/2}(1+tq)\,.
\label{ground:state:macdonald}
\ee

\bibliographystyle{ytphys}
\bibliography{refs}

\providecommand{\href}[2]{#2}\begingroup\raggedright\begin{thebibliography}{10}

\bibitem{Ruijsenaars:1986pp}
S.~N.~M. Ruijsenaars, ``{Complete Integrability of Relativistic Calogero-moser
  Systems and Elliptic Function Identities},''
  \href{http://dx.doi.org/10.1007/BF01207363}{{\em Commun. Math. Phys.}
  {\bfseries 110} (1987) 191}.

\bibitem{Ruijsenaars:1986vq}
S.~N.~M. Ruijsenaars and H.~Schneider, ``{A New Class of Integrable Systems and
  Its Relation to Solitons},''
  \href{http://dx.doi.org/10.1016/0003-4916(86)90097-7}{{\em Annals Phys.}
  {\bfseries 170} (1986) 370--405}.

\bibitem{Ruijsenaars:1988pv}
S.~N.~M. Ruijsenaars, ``{Action Angle Maps and Scattering Theory for Some
  Finite Dimensional Integrable Systems. 1. The Pure Soliton Case},''
  \href{http://dx.doi.org/10.1007/BF01238855}{{\em Commun. Math. Phys.}
  {\bfseries 115} (1988) 127--165}.

\bibitem{Gorsky:1993pe}
A.~Gorsky and N.~Nekrasov, ``{Hamiltonian systems of Calogero type and
  two-dimensional Yang-Mills theory},''
  \href{http://dx.doi.org/10.1016/0550-3213(94)90429-4}{{\em Nucl. Phys. B}
  {\bfseries 414} (1994) 213--238},
  \href{http://arxiv.org/abs/hep-th/9304047}{{\ttfamily arXiv:hep-th/9304047}}.

\bibitem{Gorsky:1993dq}
A.~Gorsky and N.~Nekrasov, ``{Relativistic Calogero-Moser model as gauged WZW
  theory},'' \href{http://dx.doi.org/10.1016/0550-3213(94)00499-5}{{\em Nucl.
  Phys. B} {\bfseries 436} (1995) 582--608},
  \href{http://arxiv.org/abs/hep-th/9401017}{{\ttfamily arXiv:hep-th/9401017}}.

\bibitem{Gorsky:1994dj}
A.~Gorsky and N.~Nekrasov, ``{Elliptic Calogero-Moser system from
  two-dimensional current algebra},''
  \href{http://arxiv.org/abs/hep-th/9401021}{{\ttfamily arXiv:hep-th/9401021}}.

\bibitem{Fock:1999ae}
V.~Fock, A.~Gorsky, N.~Nekrasov, and V.~Rubtsov, ``{Duality in integrable
  systems and gauge theories},''
  \href{http://dx.doi.org/10.1088/1126-6708/2000/07/028}{{\em JHEP} {\bfseries
  07} (2000) 028}, \href{http://arxiv.org/abs/hep-th/9906235}{{\ttfamily
  arXiv:hep-th/9906235}}.

\bibitem{Braden:1996he}
H.~W. Braden and A.~N.~W. Hone, ``{Affine Toda solitons and systems of
  Calogero-Moser type},''
  \href{http://dx.doi.org/10.1016/0370-2693(96)00499-6}{{\em Phys. Lett. B}
  {\bfseries 380} (1996) 296--302},
  \href{http://arxiv.org/abs/hep-th/9603178}{{\ttfamily arXiv:hep-th/9603178}}.

\bibitem{Braden:1997nc}
H.~W. Braden and R.~Sasaki, ``{The Ruijsenaars-Schneider model},''
  \href{http://dx.doi.org/10.1143/PTP.97.1003}{{\em Prog. Theor. Phys.}
  {\bfseries 97} (1997) 1003--1018},
  \href{http://arxiv.org/abs/hep-th/9702182}{{\ttfamily arXiv:hep-th/9702182}}.

\bibitem{Gaiotto:2009we}
D.~Gaiotto, ``{N=2 dualities},''
  \href{http://dx.doi.org/10.1007/JHEP08(2012)034}{{\em JHEP} {\bfseries 08}
  (2012) 034}, \href{http://arxiv.org/abs/0904.2715}{{\ttfamily arXiv:0904.2715
  [hep-th]}}.

\bibitem{Gaiotto:2009hg}
D.~Gaiotto, G.~W. Moore, and A.~Neitzke, ``{Wall-crossing, Hitchin Systems, and
  the WKB Approximation},''
\href{http://arxiv.org/abs/0907.3987}{{\ttfamily arXiv:0907.3987 [hep-th]}}.

\bibitem{Gaiotto:2012xa}
D.~Gaiotto, L.~Rastelli, and S.~S. Razamat, ``{Bootstrapping the superconformal
  index with surface defects},''
  \href{http://dx.doi.org/10.1007/JHEP01(2013)022}{{\em JHEP} {\bfseries 01}
  (2013) 022},
\href{http://arxiv.org/abs/1207.3577}{{\ttfamily arXiv:1207.3577 [hep-th]}}.

\bibitem{Gadde:2011uv}
A.~Gadde, L.~Rastelli, S.~S. Razamat, and W.~Yan, ``{Gauge Theories and
  Macdonald Polynomials},''
  \href{http://dx.doi.org/10.1007/s00220-012-1607-8}{{\em Commun. Math. Phys.}
  {\bfseries 319} (2013) 147--193},
  \href{http://arxiv.org/abs/1110.3740}{{\ttfamily arXiv:1110.3740 [hep-th]}}.

\bibitem{Kinney:2005ej}
J.~Kinney, J.~M. Maldacena, S.~Minwalla, and S.~Raju, ``{An Index for 4
  dimensional super conformal theories},''
  \href{http://dx.doi.org/10.1007/s00220-007-0258-7}{{\em Commun. Math. Phys.}
  {\bfseries 275} (2007) 209--254},
\href{http://arxiv.org/abs/hep-th/0510251}{{\ttfamily arXiv:hep-th/0510251
  [hep-th]}}.

\bibitem{Romelsberger:2005eg}
C.~Romelsberger, ``{Counting chiral primaries in N = 1, d=4 superconformal
  field theories},''
  \href{http://dx.doi.org/10.1016/j.nuclphysb.2006.03.037}{{\em Nucl. Phys.}
  {\bfseries B747} (2006) 329--353},
\href{http://arxiv.org/abs/hep-th/0510060}{{\ttfamily arXiv:hep-th/0510060
  [hep-th]}}.

\bibitem{Dolan:2008qi}
F.~A. Dolan and H.~Osborn, ``{Applications of the Superconformal Index for
  Protected Operators and q-Hypergeometric Identities to N=1 Dual Theories},''
  \href{http://dx.doi.org/10.1016/j.nuclphysb.2009.01.028}{{\em Nucl. Phys.}
  {\bfseries B818} (2009) 137--178},
\href{http://arxiv.org/abs/0801.4947}{{\ttfamily arXiv:0801.4947 [hep-th]}}.

\bibitem{Nazzal:2023wtw}
B.~Nazzal, A.~Nedelin, and S.~S. Razamat, ``{Ground state wavefunctions of
  elliptic relativistic integrable Hamiltonians},''
  \href{http://arxiv.org/abs/2305.09718}{{\ttfamily arXiv:2305.09718
  [hep-th]}}.

\bibitem{Hatsuda:2018lnv}
Y.~Hatsuda, A.~Sciarappa, and S.~Zakany, ``{Exact quantization conditions for
  the elliptic Ruijsenaars-Schneider model},''
  \href{http://dx.doi.org/10.1007/JHEP11(2018)118}{{\em JHEP} {\bfseries 11}
  (2018) 118}, \href{http://arxiv.org/abs/1809.10294}{{\ttfamily
  arXiv:1809.10294 [hep-th]}}.

\bibitem{Bullimore:2014upa}
M.~Bullimore and H.-C. Kim, ``{The Superconformal Index of the (2,0) Theory
  with Defects},'' \href{http://dx.doi.org/10.1007/JHEP05(2015)048}{{\em JHEP}
  {\bfseries 05} (2015) 048}, \href{http://arxiv.org/abs/1412.3872}{{\ttfamily
  arXiv:1412.3872 [hep-th]}}.

\bibitem{Bullimore:2014awa}
M.~Bullimore, H.-C. Kim, and P.~Koroteev, ``{Defects and Quantum Seiberg-Witten
  Geometry},'' \href{http://dx.doi.org/10.1007/JHEP05(2015)095}{{\em JHEP}
  {\bfseries 05} (2015) 095}, \href{http://arxiv.org/abs/1412.6081}{{\ttfamily
  arXiv:1412.6081 [hep-th]}}.

\bibitem{Nekrasov:2009rc}
N.~A. Nekrasov and S.~L. Shatashvili,
  \href{http://dx.doi.org/10.1142/9789814304634_0015}{``{Quantization of
  Integrable Systems and Four Dimensional Gauge Theories},''} in {\em
  {Proceedings, 16th International Congress on Mathematical Physics (ICMP09):
  Prague, Czech Republic, August 3-8, 2009}}, pp.~265--289.
\newblock 2009.
\newblock
\href{http://arxiv.org/abs/0908.4052}{{\ttfamily arXiv:0908.4052 [hep-th]}}.
\newblock

\bibitem{Nazzal:2018brc}
B.~Nazzal and S.~S. Razamat, ``{Surface Defects in E-String Compactifications
  and the van Diejen Model},''
  \href{http://dx.doi.org/10.3842/SIGMA.2018.036}{{\em SIGMA} {\bfseries 14}
  (2018) 036},
\href{http://arxiv.org/abs/1801.00960}{{\ttfamily arXiv:1801.00960 [hep-th]}}.

\bibitem{Nazzal:2021tiu}
B.~Nazzal, A.~Nedelin, and S.~S. Razamat, ``{Minimal $(D,D)$ conformal matter
  and generalizations of the van Diejen model},''
  \href{http://dx.doi.org/10.21468/SciPostPhys.12.4.140}{{\em SciPost Phys.}
  {\bfseries 12} no.~4, (2022) 140},
  \href{http://arxiv.org/abs/2106.08335}{{\ttfamily arXiv:2106.08335
  [hep-th]}}.

\bibitem{Alday:2010vg}
L.~F. Alday and Y.~Tachikawa, ``{Affine SL(2) conformal blocks from 4d gauge
  theories},'' \href{http://dx.doi.org/10.1007/s11005-010-0422-4}{{\em Lett.
  Math. Phys.} {\bfseries 94} (2010) 87--114},
  \href{http://arxiv.org/abs/1005.4469}{{\ttfamily arXiv:1005.4469 [hep-th]}}.

\bibitem{Kanno:2011fw}
H.~Kanno and Y.~Tachikawa, ``{Instanton counting with a surface operator and
  the chain-saw quiver},''
  \href{http://dx.doi.org/10.1007/JHEP06(2011)119}{{\em JHEP} {\bfseries 06}
  (2011) 119}, \href{http://arxiv.org/abs/1105.0357}{{\ttfamily arXiv:1105.0357
  [hep-th]}}.

\bibitem{macdonald1998symmetric}
I.~Macdonald, {\em Symmetric Functions and Hall Polynomials}.
\newblock Oxford classic texts in the physical sciences. Clarendon Press, 1998.
\newblock \url{https://books.google.it/books?id=srv90XiUbZoC}.

\bibitem{Gadde:2009kb}
A.~Gadde, E.~Pomoni, L.~Rastelli, and S.~S. Razamat, ``{S-duality and 2d
  Topological QFT},'' \href{http://dx.doi.org/10.1007/JHEP03(2010)032}{{\em
  JHEP} {\bfseries 03} (2010) 032},
  \href{http://arxiv.org/abs/0910.2225}{{\ttfamily arXiv:0910.2225 [hep-th]}}.

\bibitem{Gadde:2011ik}
A.~Gadde, L.~Rastelli, S.~S. Razamat, and W.~Yan, ``{The 4d Superconformal
  Index from q-deformed 2d Yang-Mills},''
  \href{http://dx.doi.org/10.1103/PhysRevLett.106.241602}{{\em Phys. Rev.
  Lett.} {\bfseries 106} (2011) 241602},
  \href{http://arxiv.org/abs/1104.3850}{{\ttfamily arXiv:1104.3850 [hep-th]}}.

\bibitem{Koroteev:2018isw}
P.~Koroteev, ``{A-type Quiver Varieties and ADHM Moduli Spaces},''
  \href{http://dx.doi.org/10.1007/s00220-020-03915-w}{{\em Commun. Math. Phys.}
  {\bfseries 381} no.~1, (2021) 175--207},
  \href{http://arxiv.org/abs/1805.00986}{{\ttfamily arXiv:1805.00986
  [math.AG]}}.

\bibitem{Koroteev:2019gqi}
P.~Koroteev and S.~Shakirov, ``{The quantum DELL system},''
  \href{http://dx.doi.org/10.1007/s11005-019-01247-y}{{\em Lett. Math. Phys.}
  {\bfseries 110} no.~5, (2020) 969--999},
  \href{http://arxiv.org/abs/1906.10354}{{\ttfamily arXiv:1906.10354
  [hep-th]}}.

\bibitem{Gorsky:2021wio}
A.~Gorsky, P.~Koroteev, O.~Koroteeva, and S.~Shakirov, ``{Double Inozemtsev
  limits of the quantum DELL system},''
  \href{http://dx.doi.org/10.1016/j.physletb.2022.136919}{{\em Phys. Lett. B}
  {\bfseries 826} (2022) 136919},
  \href{http://arxiv.org/abs/2110.02157}{{\ttfamily arXiv:2110.02157
  [hep-th]}}.

\bibitem{Gukov:2006jk}
S.~Gukov and E.~Witten, ``{Gauge Theory, Ramification, And The Geometric
  Langlands Program},'' \href{http://arxiv.org/abs/hep-th/0612073}{{\ttfamily
  arXiv:hep-th/0612073}}.

\bibitem{Gaiotto:2013sma}
D.~Gaiotto, S.~Gukov, and N.~Seiberg, ``{Surface Defects and Resolvents},''
  \href{http://dx.doi.org/10.1007/JHEP09(2013)070}{{\em JHEP} {\bfseries 09}
  (2013) 070}, \href{http://arxiv.org/abs/1307.2578}{{\ttfamily arXiv:1307.2578
  [hep-th]}}.

\bibitem{Kim:2016qqs}
H.-C. Kim, ``{Line defects and 5d instanton partition functions},''
  \href{http://dx.doi.org/10.1007/JHEP03(2016)199}{{\em JHEP} {\bfseries 03}
  (2016) 199}, \href{http://arxiv.org/abs/1601.06841}{{\ttfamily
  arXiv:1601.06841 [hep-th]}}.

\bibitem{Crichigno:2018adf}
P.~M. Crichigno, D.~Jain, and B.~Willett, ``{5d Partition Functions with A
  Twist},'' \href{http://dx.doi.org/10.1007/JHEP11(2018)058}{{\em JHEP}
  {\bfseries 11} (2018) 058}, \href{http://arxiv.org/abs/1808.06744}{{\ttfamily
  arXiv:1808.06744 [hep-th]}}.

\bibitem{Closset:2017zgf}
C.~Closset, H.~Kim, and B.~Willett, ``{Supersymmetric partition functions and
  the three-dimensional A-twist},''
  \href{http://dx.doi.org/10.1007/JHEP03(2017)074}{{\em JHEP} {\bfseries 03}
  (2017) 074}, \href{http://arxiv.org/abs/1701.03171}{{\ttfamily
  arXiv:1701.03171 [hep-th]}}.

\bibitem{Sciarappa:2017hds}
A.~Sciarappa, ``{Exact relativistic Toda chain eigenfunctions from Separation
  of Variables and gauge theory},''
  \href{http://dx.doi.org/10.1007/JHEP10(2017)116}{{\em JHEP} {\bfseries 10}
  (2017) 116}, \href{http://arxiv.org/abs/1706.05142}{{\ttfamily
  arXiv:1706.05142 [hep-th]}}.

\bibitem{Gadde:2020bov}
A.~Gadde, ``{Modularity of supersymmetric partition functions},''
  \href{http://dx.doi.org/10.1007/JHEP12(2021)181}{{\em JHEP} {\bfseries 12}
  (2021) 181}, \href{http://arxiv.org/abs/2004.13490}{{\ttfamily
  arXiv:2004.13490 [hep-th]}}.

\bibitem{Nekrasov:2002qd}
N.~A. Nekrasov, ``{Seiberg-Witten prepotential from instanton counting},''
  \href{http://dx.doi.org/10.4310/ATMP.2003.v7.n5.a4}{{\em Adv. Theor. Math.
  Phys.} {\bfseries 7} no.~5, (2003) 831--864},
  \href{http://arxiv.org/abs/hep-th/0206161}{{\ttfamily arXiv:hep-th/0206161}}.

\bibitem{Nekrasov:2003rj}
N.~Nekrasov and A.~Okounkov, ``{Seiberg-Witten theory and random partitions},''
  \href{http://dx.doi.org/10.1007/0-8176-4467-9_15}{{\em Prog. Math.}
  {\bfseries 244} (2006) 525--596},
  \href{http://arxiv.org/abs/hep-th/0306238}{{\ttfamily arXiv:hep-th/0306238}}.

\bibitem{Iqbal:2008ra}
A.~Iqbal, C.~Kozcaz, and K.~Shabbir, ``{Refined Topological Vertex, Cylindric
  Partitions and the U(1) Adjoint Theory},''
  \href{http://dx.doi.org/10.1016/j.nuclphysb.2010.06.010}{{\em Nucl. Phys. B}
  {\bfseries 838} (2010) 422--457},
  \href{http://arxiv.org/abs/0803.2260}{{\ttfamily arXiv:0803.2260 [hep-th]}}.

\bibitem{Kim:2011mv}
H.-C. Kim, S.~Kim, E.~Koh, K.~Lee, and S.~Lee, ``{On instantons as Kaluza-Klein
  modes of M5-branes},'' \href{http://dx.doi.org/10.1007/JHEP12(2011)031}{{\em
  JHEP} {\bfseries 12} (2011) 031},
  \href{http://arxiv.org/abs/1110.2175}{{\ttfamily arXiv:1110.2175 [hep-th]}}.

\bibitem{Kim:2012qf}
H.-C. Kim, J.~Kim, and S.~Kim, ``{Instantons on the 5-sphere and M5-branes},''
  \href{http://arxiv.org/abs/1211.0144}{{\ttfamily arXiv:1211.0144 [hep-th]}}.

\bibitem{Hosseini:2020vgl}
S.~M. Hosseini, K.~Hristov, Y.~Tachikawa, and A.~Zaffaroni, ``{Anomalies, Black
  strings and the charged Cardy formula},''
  \href{http://dx.doi.org/10.1007/JHEP09(2020)167}{{\em JHEP} {\bfseries 09}
  (2020) 167}, \href{http://arxiv.org/abs/2006.08629}{{\ttfamily
  arXiv:2006.08629 [hep-th]}}.

\bibitem{Hwang:2021xyw}
C.~Hwang, S.~S. Razamat, E.~Sabag, and M.~Sacchi, ``{Rank $Q$ E-string on
  spheres with flux},''
  \href{http://dx.doi.org/10.21468/SciPostPhys.11.2.044}{{\em SciPost Phys.}
  {\bfseries 11} no.~2, (2021) 044},
  \href{http://arxiv.org/abs/2103.09149}{{\ttfamily arXiv:2103.09149
  [hep-th]}}.

\bibitem{vanDiejen}
J.~F. van Diejen, ``Integrability of difference {C}alogero-{M}oser systems,''
  \href{https://doi.org/10.1063/1.530498}{{\em J. Math. Phys.} {\bfseries 35}
  no.~6, (1994) 2983--3004}.

\bibitem{Noumi}
M.~Noumi, S.~Ruijsenaars, and Y.~Yamada, ``The elliptic {P}ainlev\'{e} {L}ax
  equation vs. van {D}iejen's 8-coupling elliptic {H}amiltonian,''
  \href{https://doi.org/10.3842/SIGMA.2020.063}{{\em SIGMA Symmetry
  Integrability Geom. Methods Appl.} {\bfseries 16} (2020) Paper No. 063, 16}.

\bibitem{Chen:2021ivd}
J.~Chen, B.~Haghighat, H.-C. Kim, M.~Sperling, and X.~Wang, ``{E-string Quantum
  Curve},'' \href{http://arxiv.org/abs/2103.16996}{{\ttfamily arXiv:2103.16996
  [hep-th]}}.

\bibitem{Chen:2021rek}
J.~Chen, B.~Haghighat, H.-C. Kim, K.~Lee, M.~Sperling, and X.~Wang, ``{Elliptic
  quantum curves of 6d SO(N) theories},''
  \href{http://dx.doi.org/10.1007/JHEP03(2022)154}{{\em JHEP} {\bfseries 03}
  (2022) 154}, \href{http://arxiv.org/abs/2110.13487}{{\ttfamily
  arXiv:2110.13487 [hep-th]}}.

\bibitem{Razamat:2018zel}
S.~S. Razamat, ``{Flavored surface defects in 4d $\mathcal{N}=1$ SCFTs},''
  \href{http://dx.doi.org/10.1007/s11005-018-01145-9}{{\em Lett. Math. Phys.}
  {\bfseries 109} no.~6, (2019) 1377--1395},
  \href{http://arxiv.org/abs/1808.09509}{{\ttfamily arXiv:1808.09509
  [hep-th]}}.

\bibitem{Ruijsenaars:2020shk}
S.~Ruijsenaars, ``{On Razamat\textquoteright{}s $A_2$ and $A_3$ kernel
  identities},'' \href{http://dx.doi.org/10.1088/1751-8121/ab97df}{{\em J.
  Phys. A} {\bfseries 53} no.~33, (2020) 334002},
  \href{http://arxiv.org/abs/2003.11353}{{\ttfamily arXiv:2003.11353
  [math-ph]}}.

\bibitem{Seiberg:1996qx}
N.~Seiberg, ``{Nontrivial fixed points of the renormalization group in
  six-dimensions},''
  \href{http://dx.doi.org/10.1016/S0370-2693(96)01424-4}{{\em Phys. Lett. B}
  {\bfseries 390} (1997) 169--171},
  \href{http://arxiv.org/abs/hep-th/9609161}{{\ttfamily arXiv:hep-th/9609161}}.

\bibitem{Bershadsky:1997sb}
M.~Bershadsky and C.~Vafa, ``{Global anomalies and geometric engineering of
  critical theories in six-dimensions},''
  \href{http://arxiv.org/abs/hep-th/9703167}{{\ttfamily arXiv:hep-th/9703167}}.

\bibitem{Nazzal:2023bzu}
B.~Nazzal and A.~Nedelin, ``{$C_2$ generalization of the van Diejen model from
  the minimal $(D_5,D_5)$ conformal matter},''
  \href{http://arxiv.org/abs/2303.07368}{{\ttfamily arXiv:2303.07368
  [hep-th]}}.

\bibitem{Nedelin:2023syx}
A.~Nedelin, ``{Elliptic Integrable Models and Their Spectra from Superconformal
  Indices},''
\newblock 12, 2023.
\newblock \href{http://arxiv.org/abs/2312.10994}{{\ttfamily arXiv:2312.10994
  [hep-th]}}.

\bibitem{Gaiotto:2015usa}
D.~Gaiotto and S.~S. Razamat, ``{$ \mathcal{N}=1 $ theories of class $
  {\mathcal{S}}_k $},'' \href{http://dx.doi.org/10.1007/JHEP07(2015)073}{{\em
  JHEP} {\bfseries 07} (2015) 073},
  \href{http://arxiv.org/abs/1503.05159}{{\ttfamily arXiv:1503.05159
  [hep-th]}}.

\bibitem{Razamat:2013jxa}
S.~S. Razamat and M.~Yamazaki, ``{S-duality and the N=2 Lens Space Index},''
  \href{http://dx.doi.org/10.1007/JHEP10(2013)048}{{\em JHEP} {\bfseries 10}
  (2013) 048}, \href{http://arxiv.org/abs/1306.1543}{{\ttfamily arXiv:1306.1543
  [hep-th]}}.

\bibitem{Lemos:2012ph}
M.~Lemos, W.~Peelaers, and L.~Rastelli, ``{The superconformal index of class
  $S$ theories of type $D$},''
  \href{http://dx.doi.org/10.1007/JHEP05(2014)120}{{\em JHEP} {\bfseries 05}
  (2014) 120}, \href{http://arxiv.org/abs/1212.1271}{{\ttfamily arXiv:1212.1271
  [hep-th]}}.

\end{thebibliography}\endgroup

\end{document}